\documentclass{nature}
\usepackage{url,graphicx,amsmath,epstopdf,color,hyperref}

\begin{document}
\renewcommand{\arraystretch}{1.8}

\title{Origins of diamond surface noise probed by correlating single spin measurements with surface spectroscopy}

\maketitle

\author
{Sorawis Sangtawesin$^{1,*}$, Bo L. Dwyer$^{2,*}$, Srikanth Srinivasan$^{1}$, James J. Allred$^{1}$, Lila V. H. Rodgers$^{1}$, Kristiaan De~Greve$^2$, Alastair Stacey$^3$, Nikolai Dontschuk$^3$, Kane M. O'Donnell$^4$, Di Hu$^5$, D. Andrew Evans$^5$, Cherno Jaye$^6$, Daniel A. Fischer$^6$, Matthew L. Markham$^7$, Daniel J. Twitchen$^7$, Hongkun Park$^{2,8}$, Mikhail D. Lukin$^2$, Nathalie P. de Leon$^1$}

\begin{affiliations}
	\item Department of Electrical Engineering, Princeton University, Princeton, NJ 08544, USA
	\item Department of Physics, Harvard University, Cambridge, MA 02138, USA
	\item Centre for Quantum Computation and Communication Technology, School of Physics, University of Melbourne, Parkville, VIC 3010, Australia
	\item Department of Physics, Astronomy and Medical Radiation Science, Curtin University, Bentley, WA, Australia
	\item Department of Physics, Aberystwyth University, SY23 3BZ Aberystwyth, UK
	\item Material Measurement Laboratory, National Institute of Standards and Technology, Gaithersburg, MD 20899, USA
	\item Element Six, Harwell OX11 0QR, UK.
	\item Department of Chemistry and Chemical Biology, Harvard University, Cambridge, MA 02138, USA
\end{affiliations}

$\ast$ These authors contributed equally to this work

\newpage

\begin{abstract}
 
The nitrogen vacancy (NV) center in diamond exhibits spin-dependent fluorescence and long spin coherence times under ambient conditions, enabling applications in quantum information processing and sensing \cite{Maze_Nature_2008, Dobrovitski_RevCMP_2013}. NV centers near the surface can have strong interactions with external materials and spins, enabling new forms of nanoscale spectroscopy \cite{Staudacher_Science_2013, Mamin_Science_2013, Grinolds_NatNano_2014, Glenn_Nature_2018}. However, NV spin coherence degrades within 100~nanometers of the surface, suggesting that diamond surfaces are plagued with ubiquitous defects \cite{Myers_PRL_2014, Rosskopf_PRL_2014, Romach_PRL_2015, FavarodeOliveira_NatComm_2017}. Prior work on characterizing near-surface noise has primarily relied on using NV centers themselves as probes \cite{Myers_PRL_2014, Rosskopf_PRL_2014, Romach_PRL_2015, FavarodeOliveira_NatComm_2017, Lovchinsky_Science_2016, Kim_PRL_2015}; while this has the advantage of exquisite sensitivity, it provides only indirect information about the origin of the noise. Here we demonstrate that surface spectroscopy methods and single spin measurements can be used as complementary diagnostics to understand sources of noise. We find that surface morphology is crucial for realizing reproducible chemical termination, and use these insights to achieve a highly ordered, oxygen-terminated surface with suppressed noise. We observe NV centers within 10~nm of the surface with coherence times extended by an order of magnitude.

\end{abstract}

\begin{figure}
	\centering
	\includegraphics[width=0.65\columnwidth]{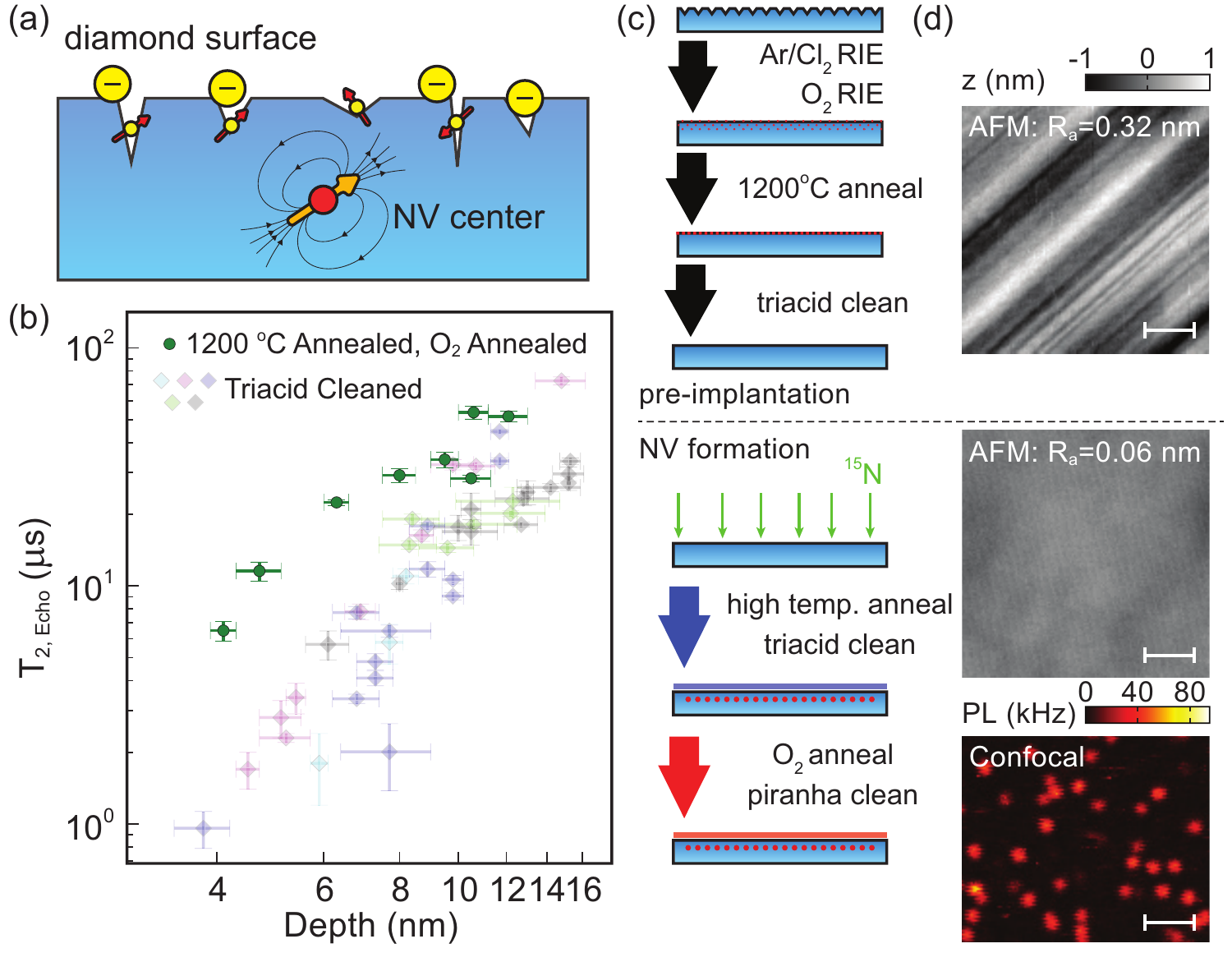}
	\caption{\textbf{Creation of shallow NV centers with long coherence times} (a) Schematic showing an NV center near the diamond surface. The surface can host defects that produce electric and magnetic field noise. (b) Hahn echo coherence time, $T_{2,\mathrm{Echo}}$, as a function of NV depth, measured across six samples with different surface conditions. Although the triacid-cleaned surfaces have different origins and processing histories (see Methods), the relationship between $T_{2,\mathrm{Echo}}$ and depth is similar across all samples, and the high-temperature- and oxygen-annealed sample exhibits significantly improved coherence times at the same depths. (c) Surface processing steps before implantation to remove surface polish damage and subsurface RIE damage, and after implantation to form NV centers and create a well-ordered oxygen surface termination. The high-temperature annealing step can be performed at 800$^\circ$C or 1200$^\circ$C. (d) AFM images of initial scaife polished diamond (top) and the final surface after oxygen annealing (middle). Scale bar is 100~nm. (bottom) Confocal image showing individually resolvable NV centers. Scale bar is 1~$\mu$m.}
	\label{fig:Fig1}
\end{figure}

Although it is easy to place NV centers near the surface by low-energy ion implantation \cite{Rosskopf_PRL_2014,Romach_PRL_2015} or delta-doping \cite{Myers_PRL_2014,Rosskopf_PRL_2014}, the surface itself can host defects that lead to noise that obscures the sensing target (Fig.~\ref{fig:Fig1}a). We observe that coherence time degrades with proximity to the surface in numerous samples with different surface conditions (Fig.~\ref{fig:Fig1}b), consistent with prior studies \cite{Myers_PRL_2014, FavarodeOliveira_NatComm_2017}, pointing to the need for new techniques to understand and control diamond surfaces. Gaining precise control over diamond surface chemistry is challenging because diamond is a chemically inert material, and also because it is hard to prepare uniform, flat diamond surfaces. Surface morphology is difficult to control because diamond's hardness makes etching and polishing non-trivial. State-of-the-art diamond polishing can achieve surface roughness below 1~nm, but the resulting surface is highly strained. Plasma etching can remove this strained layer \cite{Thoms_APL_1994, Friel_DiamRelMat_2009}, but this process is highly anisotropic and therefore small differences in initial conditions can lead to dramatic differences in final morphology and termination \cite{Howe_Carbon_2000, Wolfer_DiamRelMat_2009} (see Supplementary Information). Therefore, direct characterization of the surface is crucial for establishing that particular protocols reproducibly lead to specific, desired surface terminations.

In this work, we characterize the diamond surface by correlating photoelectron spectroscopy, X-ray absorption, atomic force microscopy (AFM), and electron diffraction with measurements of NV spin decoherence and relaxation to identify and eliminate sources of noise at the surface. We find that surface roughness leads to poor NV coherence, and we observe that surface morphology changes the density of electronic defects observed with photoelectron spectroscopy, even for the same nominal chemical termination, implying that it is critical to maintain precise control over surface purity and morphology at every processing step.

In our procedure, we remove surface and subsurface damage resulting from polishing and reactive ion etching (RIE) before ion implantation, perform high-temperature annealing to remove implantation damage, and use oxygen annealing followed by wet oxidation to terminate the surface (see Methods). In order to ensure high purity throughout processing, samples are cleaned in a refluxing mixture of concentrated perchloric, nitric, and sulfuric acids (triacid clean) before RIE and all annealing steps. Starting with scaife-polished substrates with RMS roughness of less than 1~nm, we can typically achieve final oxygen-terminated surfaces with RMS roughness $\sim$ 100~pm, as measured by AFM (Fig.~\ref{fig:Fig1}c). We show detailed examples in the Supplementary Information of contamination and irreversible surface roughening when this procedure is not followed. Using this surface processing, we extend the coherence times of NV centers within nanometers of the surface by around one order of magnitude (Fig.~\ref{fig:Fig1}b). 

To study the effects of different oxygen terminations on spin coherence, we prepare samples containing shallow NV centers using low-energy ion implantation followed by high-temperature annealing at 800$^\circ$C and triacid cleaning, and focus on detailed comparison before and after oxygen annealing. This procedure, excluding the final oxygen annealing step, is widely used for preparing shallow NV centers \cite{Staudacher_Science_2013, Mamin_Science_2013, Myers_PRL_2014, Sushkov_PRL_2014, Lovchinsky_Science_2016, Pham_PRB_2016}. We isolate the impact of the oxygen annealing step by studying the same NV centers near the surface through multiple processing cycles with and without this step.

\begin{figure}
	\centering
	\includegraphics[width=0.65\columnwidth]{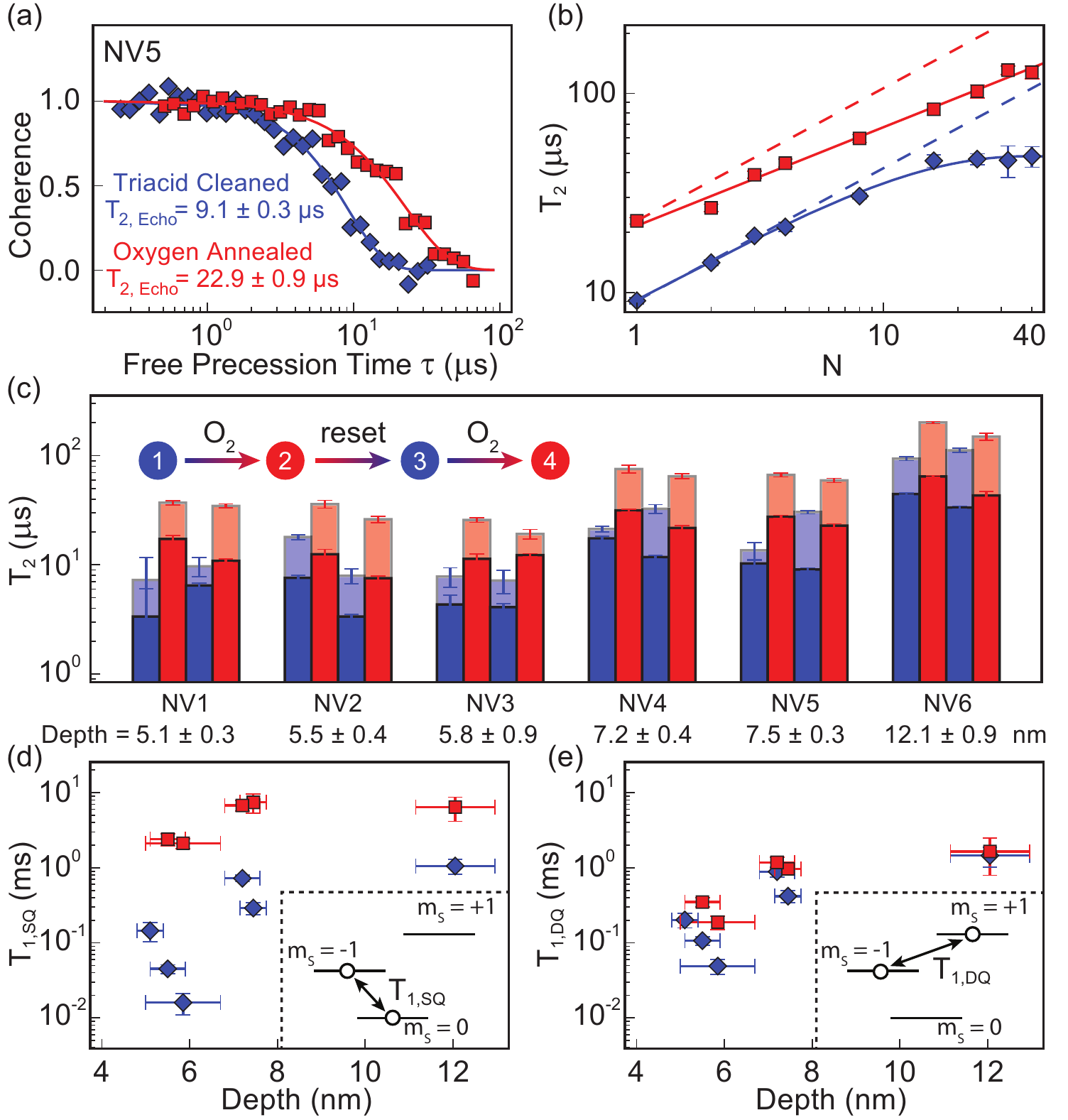}
	\caption{\textbf{Comparison of NV centers under the triacid-cleaned (red squares) and oxygen-annealed (blue diamonds) surfaces} (a) Hahn echo decay curves from the same NV center. The Hahn echo coherence time, $T_{2,\mathrm{Echo}}$, is improved after oxygen annealing. (b) Coherence time, $T_2$, as a function of number of the dynamical decoupling pulses, $N$. Dashed lines depict the scaling expected for slowly fluctuating noise, $T_2 \propto N^{2/3}$. (c) Measured $T_{2,\mathrm{Echo}}$ ($N=1$, opaque) and $T_{2, \mathrm{XY8}}$, ($N = 8$, transparent) for six NV centers through four steps of surface termination and reset (inset), showing that the oxygen annealing is reversible and reproducible. (d,e) Single- and double-quantum spin relaxation times, $T_{1,\mathrm{SQ}}$ and $T_{1,\mathrm{DQ}}$, as a function of NV depth (transitions indicated in insets). $T_{1,\mathrm{SQ}}$ shows a more pronounced improvement with oxygen annealing compared to $T_{1,\mathrm{DQ}}$, indicating that magnetic noise is more strongly suppressed than electric field noise.}
	\label{fig:Fig2}
\end{figure}

We performed single spin measurements before and after oxygen annealing. Six NV centers were randomly selected from a confocal scan, and the direct comparison of their properties under the two surfaces is shown in Fig.~\ref{fig:Fig2}. We measured the coherence time, $T_{2,\mathrm{Echo}}$, using a Hahn echo sequence for each NV center. We then studied the spectrum of the local magnetic field noise environment using XY4 and XY8 dynamical decoupling sequences \cite{deLange_Science_2010, Romach_PRL_2015}. Additionally, we studied the high frequency spectral properties of the noise using single- and double-quantum relaxation measurements for both surfaces \cite{Myers_PRL_2017}.

We observe a significant increase in $T_{2,\mathrm{Echo}}$ under the oxygen-annealed surface compared to the triacid cleaned surface (Fig.~\ref{fig:Fig2}a,c), indicating that noise at the surface is suppressed upon oxygen annealing. However, $T_{2,\mathrm{Echo}}$ still decreases as the NV centers approach the surface, indicating that surface noise remains the dominant source of decoherence. Figure~\ref{fig:Fig2}b shows an example of the measured coherence time, $T_2$, as a function of the number of $\pi$-pulses, $N$. We observe a clear improvement in $T_2$ under the oxygen-annealed surface by up to a factor of four compared to the triacid-cleaned surface for all $N$ (see Supplementary Information). For a slowly fluctuating bath, such as $^{13}$C nuclear spins or dilute P1 center electron spins, $T_2$ is expected to scale as $N^{2/3}$ (Fig.~\ref{fig:Fig2}b dashed lines) \cite{deLange_Science_2010}. We fit the data to the scaling $N^s$ and obtain $s = 0.2-0.7$ across all NV centers, and in some cases, we observe that $T_2$ saturates for $N < 40$ (Fig.~\ref{fig:Fig2}b,  Supplementary Information). This deviation from the expected $N^{2/3}$ scaling indicates that the noise at the surface is broadband, and spectral decomposition reveals a noise spectrum spanning 10~kHz to 1~MHz (see Supplementary Information).

Because the surface morphology remains smooth through oxygen annealing (Fig.~\ref{fig:Fig1}d), the termination is reversible. We demonstrate this reversibility by performing a ``surface reset'' via 800$^\circ$C vacuum annealing and triacid cleaning (Fig.~\ref{fig:Fig2}c, inset). For direct comparison between different surfaces, we measure the coherence time with Hahn echo, $T_{2,\mathrm{Echo}}$, and XY8, $T_{2,\mathrm{XY8}}$, sequences from the same NV centers across different surface terminations. After the surface reset, coherence times are reduced to values comparable to those prior to oxygen annealing. Finally, these coherence times can be restored by repeating the oxygen annealing, showing that we have reproducible control over surface termination.

We note that we achieve the longest coherence times by annealing at 1200$^\circ$C after annealing at 800$^\circ$C (Fig.~\ref{fig:Fig1}b). The higher temperature removes divacancies and multivacancy centers that form after ion implantation, which can contribute magnetic noise \cite{Twitchen_PRB_1999, Chu_NanoLett_2014, FavarodeOliveira_NatComm_2017,Yamamoto_PRB_2013}. However, we also observe that the NV center charge state is not stable after 1200$^\circ$C annealing without subsequent oxygen annealing. Therefore, to isolate the role of the oxygen termination, we have performed these experiments with only an 800$^\circ$C post-implantation anneal.


We probe the separate contributions of electric and magnetic noise by measuring relaxation rates between different levels in the NV ground state (Fig.~\ref{fig:Fig2}d,e). The single-quantum (SQ) transition can be driven by magnetic noise, while the double-quantum (DQ) transition is magnetically forbidden and can thus be used to probe electric field noise. Comparison of the measured SQ and DQ spin relaxation times allows for the extraction of SQ and DQ transition rates, which are a reflection of the relative contributions of electric and magnetic noise \cite{Myers_PRL_2017}. Figure~\ref{fig:Fig2}d shows the measured SQ and DQ spin relaxation times, $T_{1,\mathrm{SQ}}$ and $T_{1,\mathrm{DQ}}$, measured at $B_z =$ 40 G for the two different surface terminations. We observe an improvement in $T_{1,\mathrm{SQ}}$ of 1-2 orders of magnitude after oxygen annealing, indicating that the high frequency magnetic field noise is strongly suppressed, consistent with $T_2$ measurements that probe the magnetic field noise at lower frequencies. In comparison, $T_{1,\mathrm{DQ}}$ exhibits a small improvement (less than a factor of three) after oxygen annealing (Fig.~\ref{fig:Fig2}e). Dynamical decoupling, SQ, and DQ relaxation measurements are sensitive to different frequency regimes, but the DQ transition rate is expected to scale inversely with frequency \cite{Myers_PRL_2017}, allowing for extrapolation to other frequencies. Spectral comparison of the dynamical decoupling data and DQ relaxation data indicates that the electric field noise is not the dominant source that limits the coherence of NV centers under either surface termination (see Supplementary Information).

\begin{figure}
	\centering
	\includegraphics[width=0.65\columnwidth]{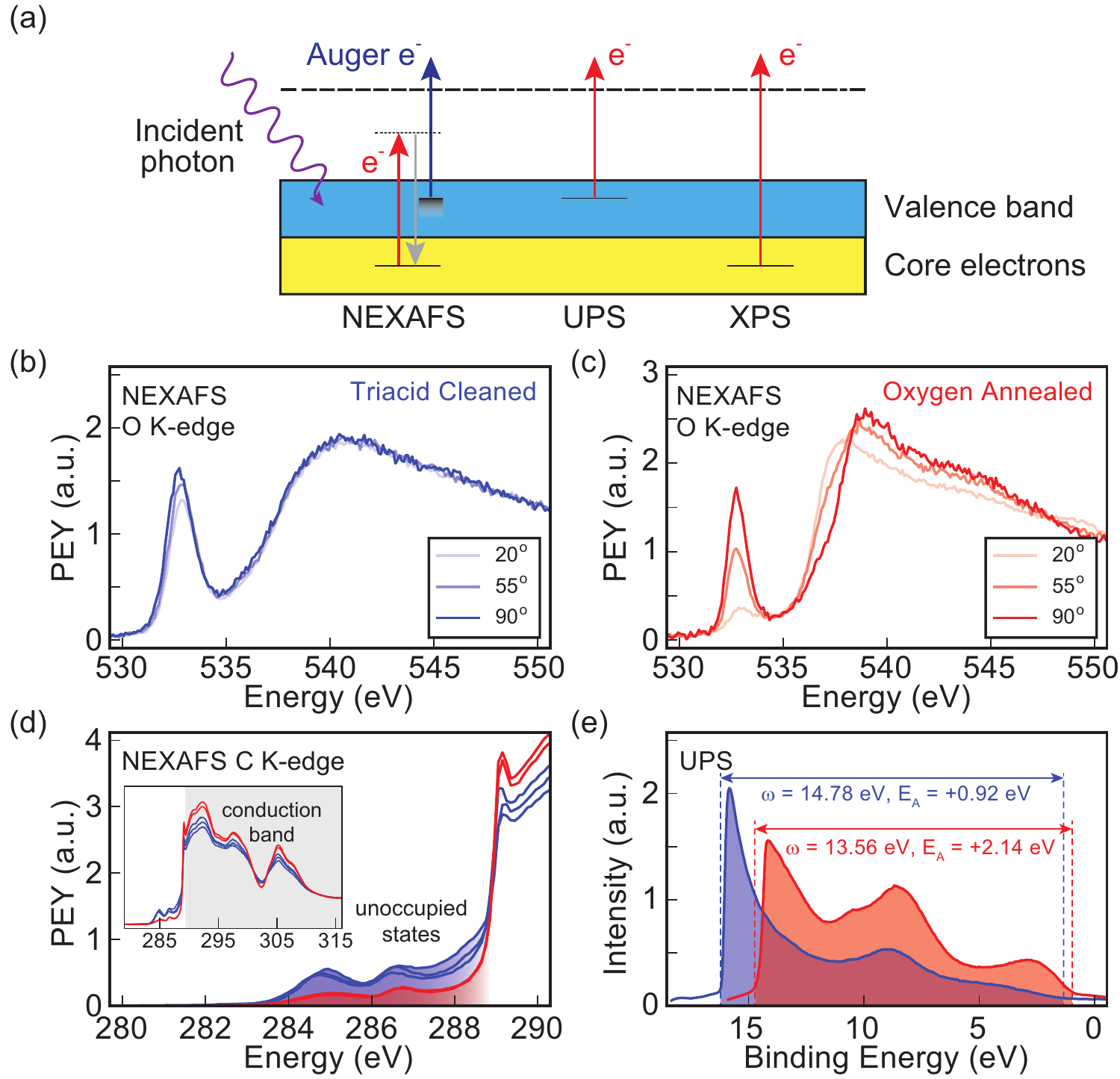}
	\caption{\textbf{Surface spectroscopy comparison between triacid-cleaned (blue) and oxygen-annealed (red) surfaces} (a) Schematic showing surface-sensitive spectroscopy techniques used to probe the diamond surface. (b,c) Polarization dependence of NEXAFS spectra at the oxygen K-edge. The polarization angle is defined relative to the surface normal. The y-axis for all NEXAFS spectra is the partial electron yield (PEY), normalized and baseline subtracted (see Methods). The oxygen-annealed surface spectrum exhibits strong polarization dependence, indicating that oxygen groups at the surface are well-ordered. (d) NEXAFS spectra at the carbon K-edge from three triacid-cleaned samples (blue curves) and two oxygen-annealed samples (red curves). Triacid-cleaned surfaces show a higher density of unoccupied states (shaded regions). The full energy range is shown in the inset. (e) UPS spectra with excitation energy $h\nu =$ 21.2~eV of the triacid-cleaned (blue) and oxygen-annealed (red) surfaces. The spectral width, $\omega$, is indicated by the shaded regions. The oxygen-annealed surface exhibits a higher positive electron affinity, $E_A$, (2.14~eV) than the triacid-cleaned surface (0.92~eV).}
	\label{fig:Fig3}
\end{figure}

NV-based measurements indirectly suggest that the triacid-cleaned and oxygen-annealed surfaces have different electronic structure. To directly characterize the structure and chemical composition of the two different oxygen-terminated surfaces, we employ a variety of surface-sensitive spectroscopy techniques (Fig.~\ref{fig:Fig3}a). Near-edge X-ray absorption fine structure spectroscopy (NEXAFS) probes the density of unoccupied states near the surface, ultraviolet photoelectron spectroscopy (UPS) gives information about the Fermi energy and electron affinity, and X-ray photoelectron spectroscopy (XPS) yields information about the chemical state of the surface termination.

\begin{figure}
	\centering
	\includegraphics[width=0.65\columnwidth]{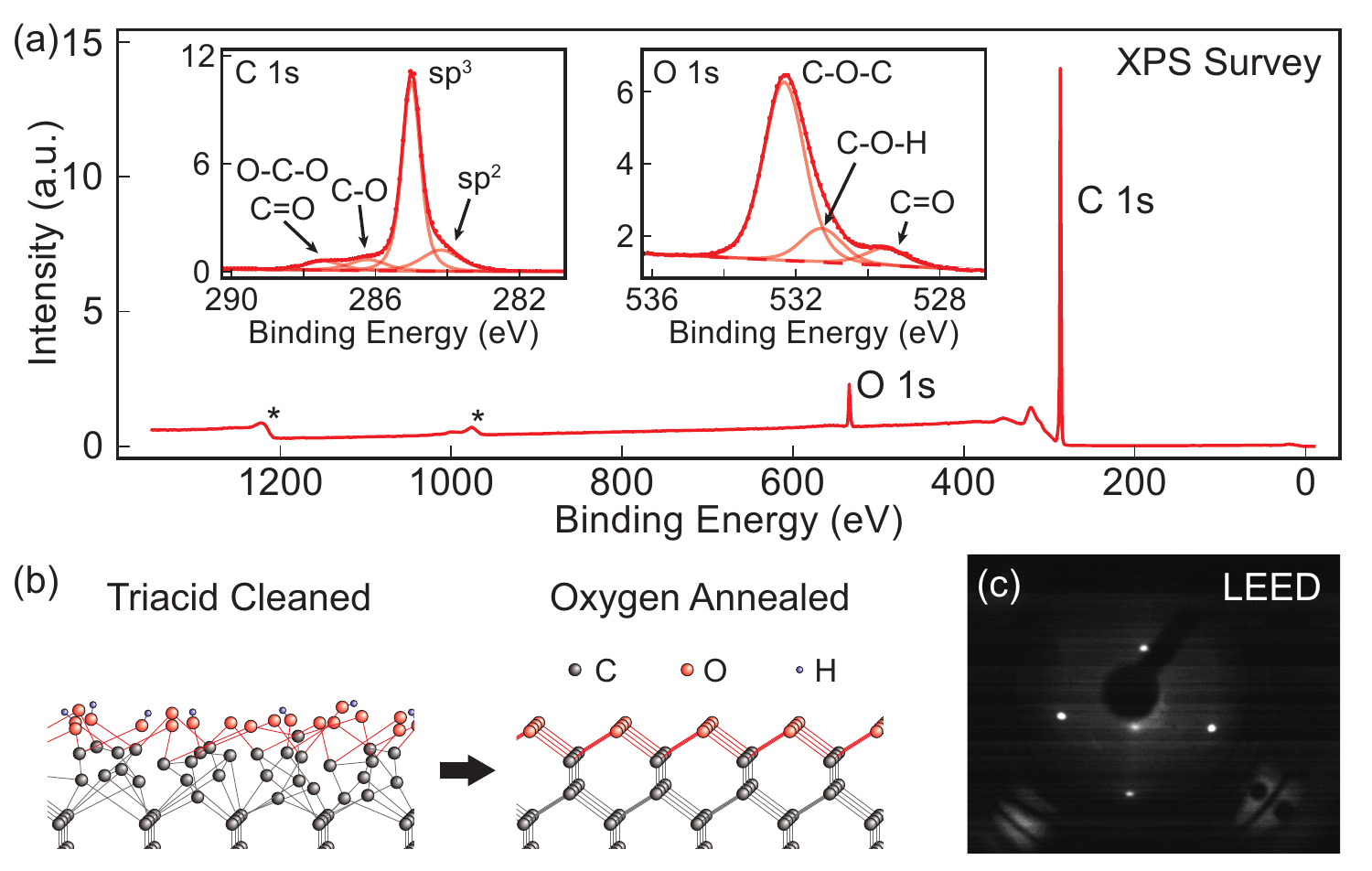}
	\caption{\textbf{Detailed characterization of the oxygen-annealed surface} (a) XPS survey scan showing dominant carbon and oxygen peaks. Peaks marked with ($\ast$) are Auger peaks associated with carbon and oxygen. (left inset) High-resolutionigh-resolution XPS carbon $1s$ spectrum showing a dominant diamond $sp^3$ peak with three side peaks assigned to $sp^2$ carbon, and carbon singly- and doubly-bonded to oxygen. (right inset) High-resolutionigh-resolution XPS oxygen $1s$ spectrum showing a dominant ether (C-O-C) peak and two satellite peaks assigned to alcohol (C-O-H) and ketone (C=O). (b) Ball and stick diagrams illustrating the surface termination before and after oxygen annealing. The disordered, acid-cleaned surface hosts a mixture of groups, which transform into a highly-ordered, predominantly ether-terminated surface upon oxygen annealing. (c) LEED pattern showing a $1\times1$ reconstruction of the oxygen-annealed surface, consistent with ether termination.}
	\label{fig:Fig4}
\end{figure}

The NEXAFS spectra at the oxygen K-edge for the two surfaces are qualitatively similar (Fig.~\ref{fig:Fig3}b,c). Both exhibit a sharp $\pi^{\ast}$ peak at 532.5~eV and a broad $\sigma^{\ast}$ shoulder at around 540~eV, indicating similar chemical states. Varying the angle of incidence of the linearly-polarized X-rays changes the relative polarization with respect to the surface normal. As this angle is varied, the signal changes dramatically for the oxygen-annealed surface (Fig.~\ref{fig:Fig3}c), while the triacid-cleaned surface shows no variation (Fig.~\ref{fig:Fig3}b). Strong polarization dependence arises from distinct and well-resolved bond orientations \cite{Stohr_Springer_1992}, indicating that the oxygen-annealed surface is highly ordered at the atomic scale, while the oxygen groups in the acid-cleaned surface are disordered (Fig.~\ref{fig:Fig4}b).

The NEXAFS spectra at the carbon K-edge (Fig.~\ref{fig:Fig3}d) show a characteristic exciton peak at 289.2~eV and a second absolute band gap at 302.2~eV \cite{Shpilman_DiamRelMat_2014}. At energies below the exciton peak at the conduction band edge, both surfaces exhibit two peaks, one at 285~eV that is assigned to $sp^2$ carbon, and one at 286.5~eV associated with oxygen termination \cite{Shpilman_DiamRelMat_2014}. However, the triacid-cleaned surface has an average of 2.4~times higher density of unoccupied states below the conduction band edge, indicated by the area under the pre-edge region. These energetically deep unoccupied states at the surface can potentially act as electronic traps that host unpaired electrons, which can contribute both magnetic and electric field noise \cite{Stacey_arXiv_2018}. Furthermore, a morphologically rough surface after the same surface preparation and oxygen annealing exhibits a much larger density of unoccupied states than the smooth surface (see Supplementary Information). Using UPS, we observe that the oxygen-annealed surface exhibits a positive electron affinity of 2.14~eV, compared to 0.92~eV for the triacid-cleaned surface (Fig.~\ref{fig:Fig3}e), indicating that the two surfaces possess drastically different electronic structure. To the best of our knowledge, this electron affinity is the largest reported for oxygen-terminated diamond \cite{Maier_PRB_2001}.

Combining the data from surface spectroscopy and NV measurements, we conclude that disorder at the surface can lead to unoccupied defect states near the conduction band edge of diamond, which in turn lead to rapid decoherence of NV centers near the surface. These defect states give rise to broadband magnetic noise that cannot be circumvented by simple dynamical decoupling. It is therefore important for future applications in nanoscale sensing to devise methods to eliminate disorder and defect states at the diamond surface.

We now turn our attention to the chemical identification of the well-ordered, oxygen-annealed surface. XPS (Fig.~\ref{fig:Fig4}a) reveals that the only detectable atoms are carbon and oxygen. The oxygen peak comprises 6--7\% of the signal, corresponding to approximately monolayer surface coverage (see Supplementary Information). High-resolution XPS was used to probe the structure of the carbon and oxygen $1s$ peaks in detail (Fig.~\ref{fig:Fig4}a, inset). The carbon $1s$ spectrum shows a dominant peak at 285~eV, which we assign to diamond $sp^3$ carbon. Two satellite peaks at higher binding energies of +1.2~eV and +2.4~eV correspond to carbon singly- and doubly-bonded to oxygen, respectively \cite{Kitagawa_IET_2011}. The peak at lower binding energy of -0.8~eV is assigned to $sp^2$ carbon. The oxygen $1s$ spectrum shows a major peak at 532.3~eV and two satellite peaks at lower binding energies of -1.0~eV and -2.8~eV with a relative ratio of 10:1:1. These peaks have been previously assigned to ether, alcohol, and ketone, respectively \cite{Baldwin_PRB_2014}. Low-energy electron diffraction (LEED) indicates that the surface is $1\times1$ reconstructed (Fig.~\ref{fig:Fig4}c). Combining the XPS and LEED data, we assign the surface as predominantly ether-terminated ($\sim$90\%), with a minority mixture of alcohol and ketone groups (Fig.~\ref{fig:Fig4}b). Additionally, our measured electron affinity is consistent with density functional theory calculations of an ether-terminated surface \cite{Sque_PRB_2006}. We note that all of the surface spectroscopy techniques are unable to directly detect hydrogen, and we thus cannot exclude that the mixed surface includes residual hydrogen, although the large positive electron affinity rules out significant hydrogen incorporation \cite{Maier_PRB_2001}.

While the coherence of shallow NV centers is significantly improved by the techniques presented here, these coherence times remain far from typical bulk values \cite{Balasubramanian_NatMat_2009}. The present work suggests a number of promising avenues for future study. The $sp^2$ peak observed in NEXAFS and XPS is a deep electronic trap, and is a natural target for improvement \cite{Stacey_arXiv_2018}. It was also recently demonstrated that NV center coherence can be improved by implanting nitrogen through a boron-doped layer \cite{FavarodeOliveira_NatComm_2017}. Combining such strategies with our surface preparation could yield even longer spin coherence.  Finally, it is unknown what contribution adventitious carbon contamination makes to magnetic and electric field noise. Our on-going work includes preparing and interrogating surfaces in ultrahigh vacuum conditions to disentangle the contributions of chemical surface termination and exogenous contamination. 

Our approach of combining surface spectroscopy with single spin measurements can be applied to the future development of novel surface terminations. We have shown that surface morphology and electronic structure measurements can help to evaluate which surfaces are likely to lead to further improvements in NV coherence, which can provide useful benchmarking for rapid exploration of new surface chemistry. More broadly, the strategy of correlating surface spectroscopy with qubit measurements can be applied to a variety of quantum platforms that also exhibit deleterious effects from surfaces and interfaces, such as superconducting qubits \cite{Gao_APL_2008}, trapped ions \cite{Hite_MRS_2013}, and shallow donors \cite{Paik_PRB_2010}.

\bibliography{Oxygen_Surfaces} 

\newpage
\begin{methods}

\subsection{Samples} \label{sec:samples}

In this letter, we present NV coherence data from several samples in Fig.~\ref{fig:Fig1}b. Here, we describe the different samples:
\begin{itemize}
	\item Sample~A: The sample used for data in Fig.~\ref{fig:Fig2}. Commercially available electronic grade diamond (Element Six) with $<$5 ppb nitrogen and $<$1 ppb boron. 
	\item Sample~B: The sample that was subjected to 1200$^\circ$C annealing followed by oxygen annealing, presented in Fig.~\ref{fig:Fig1}b. This sample originated from the same crystal as sample A. The crystal was sliced prior to any processing described in Fig.~\ref{fig:Fig1}c. 
	\item Sample~C: Another sample that originated from the same crystal as sample~A. The crystal was sliced prior to any processing described in Fig.~\ref{fig:Fig1}c. 
	\item Sample~D: An electronic grade sample with $^{12}$C-enriched layer that was processed according to Fig.~\ref{fig:Fig1}c. 
	\item Sample~E: An electronic grade sample with $^{12}$C-enriched layer with rough, as-grown surface that was subsequently processed with Ar/Cl$_2$ and O$_2$ RIE. 
	\item Sample~F: An electronic grade sample with $^{12}$C-enriched layer. The surface is left as-grown. 
\end{itemize}

In addition to these NV samples, surface spectroscopy data is presented from several electronic grade samples. Three triacid-cleaned samples and three oxygen-high-resolutionannealed samples, one of which is morphologically rough, were used for NEXAFS and high-resolution XPS (Fig.~\ref{fig:Fig3}b--d, Fig.~\ref{fig:Fig4}a, inset). A boron doped sample (0.1 ppm boron and $<$5 ppb nitrogen) was used for UPS spectroscopy (Fig.~\ref{fig:Fig3}e) to prevent charging. Finally, a lower purity sample ($<$1 ppm nitrogen and $<$0.5 ppm boron, Element Six "standard grade") was used for oxygen annealing calibration.


\subsection{Sample Preparation}
Our method for preparing a high quality diamond surface prior to ion implantation relies on a multi-step process to remove surface and subsurface damage. Unless indicated otherwise, all samples described above are laser cut and scaife-polished to a RMS roughness of less than 1~nm with a $(100)$ major face and $\langle110\rangle$ edges, specified to within 3$^\circ$. In order to prepare substrates for implantation, reactive ion etching was performed using an inductively coupled plasma (ICP) with the following parameters: 400~W ICP power, 250~W substrate bias RF power, 25~sccm Ar, 40~sccm Cl$_2$, 8~mTorr for 30~minutes followed by 700~W ICP, 100~W substrate bias, 30~sccm O$_2$, 10~mTorr for 25~minutes (Plasma-Therm Versaline ICP RIE). These two RIE steps etch approximately 2~$\mathrm{\mu m}$ and 4~$\mathrm{\mu m}$ of the subsurface polish damage layer, respectively \cite{Friel_DiamRelMat_2009}.

In order to remove residual subsurface damage from ICP RIE, 1200$^\circ$C vacuum annealing is performed in a Lindberg Blue tube furnace with high purity ($>$99.5\%) alumina ceramic tubes at pressures between $1\times 10^{-7}$~Torr and $2\times 10^{-6}$~Torr with the following sequence, starting from room temperature:

\begin{enumerate}
	\item Ramp to 100$^\circ$C over 1~hour. Hold for 11~hours.
	\item Ramp to 400$^\circ$C over 4~hours. Hold for 8~hours.
	\item Ramp to 800$^\circ$C over 6--12~hours. Hold for 8~hours.
	\item Ramp to 1200$^\circ$C over 6--12~hours. Hold for 2~hours.
	\item Let cool to room temperature.
\end{enumerate}

This annealing results in a $<$3~nm layer of amorphous carbon at the surface, which is subsequently removed by cleaning the sample in a refluxing 1:1:1 mixture of concentrated sulfuric, nitric, and perchloric acids (triacid clean) for at least one hour. The conversion of material to amorphous carbon and subsequent removal is critical for removing subsurface damage resulting from RIE processing. Annealing at lower pressures (below $1\times 10^{-7}$~Torr) does not result in a thick layer of amorphous carbon, and thus does not remove this damage layer (see Supplementary Information).

Sample A was then sent for $^{15}$N ion implantation (Innovion) with the recipe: dose $= 1\times 10^9\ \mathrm{cm^{-2}}$, energy $=$ 3~keV, and 0$^\circ$ tilt. Other samples were implanted with the same parameters, except with doses of $3\times 10^9\ \mathrm{cm^{-2}}$ for samples~B and F and $5\times 10^8\ \mathrm{cm^{-2}}$ for samples~C, D, and E. Following implantation, all samples are triacid cleaned and 800$^\circ$C annealed in vacuum using the same recipe as above, with or without the 1200$^\circ$C step. Another triacid clean following this vacuum anneal results in the condition referred to as the ``triacid cleaned" surface throughout the text.

To create the oxygen-terminated surface, the sample is then annealed at 445--450$^\circ$C in a tube furnace (Lindberg Blue Mini-Mite with high-purity quartz process tube) under continuous flow of O$_2$ at atmospheric pressure for 4~hours. The oxygen flow is regulated with a mass flow controller, and the outlet of the process tube is connected to a bubbler to prevent backflow of gases. The input gases, oxygen (for annealing) and nitrogen (for venting the furnace), are filtered via SAES Sentrol point-of-use purifiers, MC1-203F and MC1-902F, respectively. Following the oxygen anneal, the sample is cleaned in a 1:2 mixture of hydrogen peroxide in concentrated sulfuric acid (``piranha"). The resulting sample condition is referred to as the ``oxygen-annealed" surface throughout the text. Finally, Sample~B was annealed at 1200$^\circ$C and subsequently oxygen annealed to achieve the best spin coherence times, shown in Fig.~\ref{fig:Fig1}b.

XPS is used between each step to verify that the surface is contamination free at the $0.1\%$ level, which is the sensitivity limit of the instrument. If any heteroatoms other than C and O are found (e.g. Na, Cl, Si), the sample is repeatedly cleaned with either triacid or piranha until the contamination is eliminated. In the Supplementary Information, we show  examples of XPS spectra from a contaminated sample before and after acid cleaning, as well as micromasking and surface roughening that can result from contamination.

\subsection{Process Calibration for Oxygen Annealing}
Since diamond etches when heated in an oxygen atmosphere~\cite{Gaebel_DiamRelMat_2012}, changing the surface termination while avoiding etching requires careful temperature calibration. Our process proceeds as follows:

\begin{enumerate}
	\item Clean sample in 1:2 hydrogen peroxide in sulfuric acid (piranha). Verify that the sample is contaminant-free in XPS.
	\item In the furnace, ramp to the target temperature over 4 hours, and  anneal at the target temperature for another 4 hours.
	\item Repeat step 1.
	\item Examine the sample in AFM to check for morphological changes at the surface.
	\item Cycle back to first step and increase target temperature.
\end{enumerate}

Previous studies showed that diamond starts to etch in oxygen around 500$^\circ$C ~\cite{Gaebel_DiamRelMat_2012}. Therefore, we begin with 450$^\circ$C for our calibration and choose the final temperature to be the highest temperature that does not produce pitting on the sample. Examples of AFM images taken after annealing at different temperatures are shown in the Supplementary Information.

\subsection{XPS and AFM Characterization}
XPS survey scans in Fig.~\ref{fig:Fig4}a and AFM images in Fig.~\ref{fig:Fig1}c were performed at the Imaging and Analysis Center (IAC) at Princeton University. XPS was performed with a Thermo Fisher K-Alpha spectrometer, collecting photoelectrons normal to the surface. AFM was performed interchangeably with either a Bruker Nanoman or a Bruker ICON3 AFM operating in AC tapping mode (AFM tip:  Asylum Research AC160TS-R3, resonance frequency 300~kHz). Each diamond was thoroughly cleaned with either triacid or piranha before AFM scans were performed. Large-scale ($5\times 5$~$\mu$m) and small-scale ($1\times 1$~$\mu$m or $0.5\times 0.5$~$\mu$m) scans were performed in several distinct areas of the diamonds, away from the edge. In general, for the same sample, no clear variation of the RMS roughness was observed across the interrogated areas.

\subsection{NEXAFS and High-Resolution XPS}
In near-edge X-ray absorption fine structure spectroscopy (NEXAFS), monochromatic X-rays excite core electrons, and secondary electron yield is measured as a function of the incident X-ray energy, giving a signal that is proportional to the density of unoccupied states near the surface.  In XPS, incident X-rays ionize core electrons, and the measured binding energy is sensitive to the chemical environment of the ionized atom.

Unless indicated otherwise, all NEXAFS data and high-resolution XPS spectra (Fig.~\ref{fig:Fig4} insets) in the main text and Supplementary Information were acquired at the Australian Synchrotron soft X-ray spectroscopy beamline, using light from an APPLE II undulator generating linearly polarized photons and passed through a plane-grating monochromator. Prior to scanning, the samples were annealed \textit{in situ} at at 430$^\circ$C to remove adventitious carbon \cite{Barr_JVacSci_1995}. 

Carbon K-edge and oxygen K-edge NEXAFS were collected in partial electron yield mode with grid biases of 220~V and 440~V, respectively. The spectra are processed and calibrated by first dividing by the total incident power measured using photoelectrons from clean gold foil in the chamber, subtracting the average pre-edge background (270--275~eV for carbon, 520--525~eV for oxygen), and normalizing to the post-edge electron yield (315--320~eV for carbon, 558--560~eV for oxygen). The energy is calibrated by setting the sharp $\sigma$* exciton peak of the gold foil to 291.65~eV \cite{Watts_JElecSpec_2008}.

High-resolution XPS spectra were analyzed using a SPECS Phoibos 150 hemispherical analyzer with the pass energy set to 5~eV, resulting in a linewidth of better than 0.1~eV. An excitation photon energy of 600~eV  was used. XPS spectra were fitted using CasaXPS. A linear fit to the pre-edge was first subtracted from the data to account for the rising secondary electron tail apparent in spectra acquired with a photon energy close to the core level energy. Subsequently, a universal Tougaard background was subtracted and Voigt functions were used to fit the resulting spectra. Each component function was constrained to have the same FWHM as all others within the same spectrum. We find that the carbon signal fit residual is minimized by fitting two side peaks on the high binding energy side rather than one, and that it does not improve by fitting three side peaks. We identify these two peaks as carbon singly- and doubly-bonded to oxygen.

\subsection{Additional XPS, UPS, and LEED Measurements}
Additional XPS (Supplementary Information), UPS, and LEED measurements were carried out in a custom UHV spectrometer at Aberystwyth University. X-ray excitation was provided by a VG twin-anode (Mg and Al) source and He I UV radiation was provided by a SPECS UVS 300 source. Photoelectrons were collected at normal emission by a SPECS Phoibos 100 analyzer using a 2D CCD electron detector. 

In UPS, ultraviolet photons (21.2~eV) ionize valence electrons, and their binding energy can then be used to determine the Fermi energy and electron affinity. For XPS, the sample was kept at earth potential while for UPS, a bias of $-2$~V was applied to the sample to enable collection of low energy electrons over a range of sample work functions. The electron analyzer was operated in wide angle mode to sample band edge states averaged in momentum space. Since the apparent binding energy of electron states measured by photoelectron spectroscopy is affected by surface charging and photovoltage generation \cite{Williams_APL_2014}, we calibrate the valence band edge against the Fermi edge of a tantalum standard. Rear-view VG LEED optics were used to record surface electron diffraction patterns. The beam energy was set to 86~eV for the diffraction pattern shown in Fig.~\ref{fig:Fig4}c.

\subsection{NV Measurement Setup}
NV measurements were performed on a home-built confocal microscope. NV centers are excited by a 532 nm optically pumped solid state laser (Coherent Sapphire LP 532-300), which is modulated with an acousto optic modulator (Isomet 1205C-1). The beam is scanned using galvo mirrors (Thorlabs GVS012) and projected into an oil immersion objective (Nikon, Plan Fluor 100$\times$, $\mathrm{NA} = 1.30$) with a telescope in a $4f$ configuration. Laser power at the back of the objective was kept between 60--100 $\mu$W, approximately $25\%$ of the saturation power of a single NV center, in order to avoid irreversible photobleaching. A dichroic beamsplitter (Thorlabs DMLP567) separates the excitation and collection pathways, and fluorescence is measured using a fiber-coupled avalanche photodiode (Excelitas SPCM-AQRH-44-FC). A neodymium magnet is used to introduce a DC magnetic field for Zeeman splitting, and the orientation of the magnetic field was aligned to within 1$^\circ$ of the NV axis using a combination of a rotation stage and a goniometer.

Spin manipulation on the NV center was accomplished using microwaves generated by a dual-channel signal generator (R\&S SMATE200A). The two channels are independently gated with fast SPDT switches (Mini-Circuits ZASWA-2-50DR+) and combined with a resistive combiner (Mini-Circuits ZFRSC-42-S+) for double-quantum measurement capability. The combined signal is then amplified with a high-power amplifier (Mini-Circuits ZHL-16W-43+) and delivered to the sample via a coplanar stripline. The stripline is fabricated by depositing 10 nm Ti, 1000 nm Cu, and 200 nm Au on a microscope coverslip. Following metallization, the stripline is photolithographically defined and etched with gold etchant and hydrofluoric acid. Finally, a 100 nm layer of Al$_2$O$_3$ is deposited on top of the fabricated stripline via atomic layer deposition (ALD) to protect the metal layer. This Al$_2$O$_3$ layer is crucial for separating the diamond surface from the metal layer of the stripline, which can contaminate the diamond surface (see Supplementary Information). Pulse timing is controlled with a Spincore PulseBlaster ESR-PRO500 with 2~ns timing resolution, and phase control of the NV microwave pulses is achieved with an arbitrary waveform generator (Agilent 33622A). Hahn echo experiments were performed at $B = 1900~\mathrm{G}$ to average out the free precession of the 1.1\% natural abundance of $^{13}$C in the sample. 

To avoid effects of pulse errors during dynamical decoupling, we alternate the phase of each $\pi$-pulse using the XY4 protocol for the 4-pulse sequence, the XY8 protocol for the 8-pulse sequence, and repeated XY8 protocol for higher order sequences \cite{Wang_PRB_2012}. 

NV depths were measured by placing a drop of microscope immersion oil (Nikon NF2) on the surface and measuring the proton NMR signal arising from the oil \cite{Pham_PRB_2016}.

\end{methods}

\begin{addendum}
	\item We thank Adam Gali, Joseph Tabeling, and Jim Butler for numerous discussions about diamond surfaces, Nan Yao, Yao-Wen Yeh, and John Schreiber at the Princeton Imaging and Analysis Center for help with diamond surface characterization, Hans Bechtel and David Kilcoyne at the Advanced Light Source as well as Arthur Woll at the Cornell High Energy Synchrotron Source for advice about surface spectroscopy techniques, and Jeff Thompson for other fruitful discussions. This work was supported by the NSF under the CAREER program (grant DMR-1752047) and through the Princeton Center for Complex Materials, a Materials Research Science and Engineering Center (grant DMR-1420541). JJA acknowledges support from the National Science Foundation Graduate Research Fellowship Program, and LVHR acknowledges support from the Department of Defense through the National Defense Science and Engineering Graduate Fellowship Program. AS and ND acknowledge support from the Australian Research Council (CE170100012). Part of this research was undertaken on the Soft X-ray spectroscopy beamline at the Australian Synchrotron, part of ANSTO.
	
	\item[Competing Interests] The authors declare that they have no
	competing financial interests.
	\item[Correspondence] Correspondence and requests for materials
	should be addressed to npdeleon@princeton.edu.
\end{addendum}
\end{document}


\renewcommand{\arraystretch}{1.8}
\renewcommand{\thefigure}{S\arabic{figure}}
\renewcommand{\theequation}{S\arabic{equation}}

\title{Supplementary Information: Origins of diamond surface noise probed by correlating single spin measurements with surface spectroscopy}

\author{Sorawis Sangtawesin$^{1,*}$, Bo L. Dwyer$^{2,*}$, Srikanth Srinivasan$^{1}$, James J. Allred$^{1}$, Lila V. H. Rodgers$^{1}$, Kristiaan De~Greve$^2$, Alastair Stacey$^3$, Nikolai Dontschuk$^3$, Kane M. O'Donnell$^4$, Di Hu$^5$, D. Andrew Evans$^5$, Cherno Jaye$^6$, Daniel A. Fischer$^6$, Matthew L. Markham$^7$, Daniel J. Twitchen$^7$, Hongkun Park$^{2,8}$, Mikhail D. Lukin$^2$, Nathalie P. de Leon$^1$}

\email{npdeleon@princeton.edu}

\affiliation{$^1$Department of Electrical Engineering, Princeton University, Princeton, NJ 08544, USA}
\affiliation{$^2$Department of Physics, Harvard University, Cambridge, MA 02138, USA}
\affiliation{$^3$Centre for Quantum Computation and Communication Technology, School of Physics, University of Melbourne, Parkville, VIC 3010, Australia}
\affiliation{$^4$Department of Physics, Astronomy and Medical Radiation Science, Curtin University, Bentley, WA, Australia}
\affiliation{$^5$Department of Physics, Aberystwyth University, SY23 3BZ Aberystwyth, UK}
\affiliation{$^6$Material Measurement Laboratory, National Institute of Standards and Technology, Gaithersburg, MD 20899, USA}
\affiliation{$^7$Element Six, Harwell OX11 0QR, UK}
\affiliation{$^8$Department of Chemistry and Chemical Biology, Harvard University, Cambridge, MA 02138, USA}

\maketitle

\tableofcontents
oxygen-annealed
\section{Effects of Sample Contamination}
\begin{figure}[h!]
	\centering
	\includegraphics[width=0.7\columnwidth]{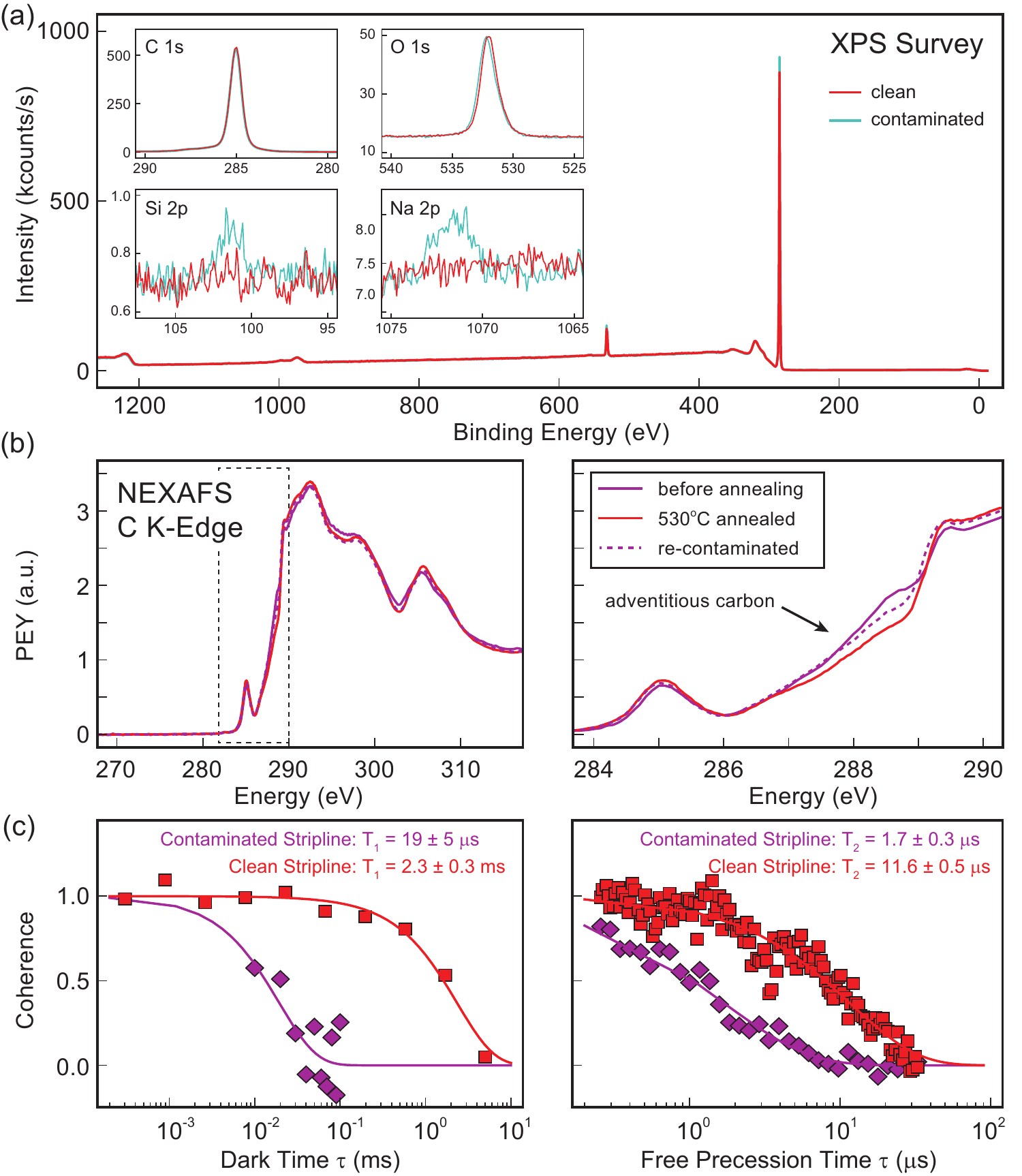}
	\caption{(a) XPS spectra from a diamond sample with Si and Na contamination before (cyan) and after (red) acid cleaning. The main panel shows the full energy range survey scans, while insets depict higher resolution peak scans at the C 1$s$, O 1$s$, Si 2$p$, and Na 2$p$ transitions. (b) Normalized NEXAFS carbon K-edge spectra and a close-up of the pre-edge region of the same sample before and after \textit{in situ} annealing at 530$^\circ$C, showing the removal of pre-edge states attributed to adventitious carbon. Dashed line shows the spectrum after re-exposure to atmosphere. (c) Single-quantum relaxation and Hahn echo coherence, measured from the same NV with a contaminated stripline in which the ALD layer has been scratched, compared with a new stripline.}
	\label{fig:SI_Contamination}
\end{figure}

In this work, we emphasize that prior to any surface processing step, it is critical to start with a morphologically smooth and contamination-free surface, since impurities can cause irreversible surface damage through processing. It is also important to monitor the surface roughness and contamination between each step. To ensure purity between each processing step, we clean the diamonds in triacid or piranha solution and check for contaminants in XPS before proceeding. Figure~\ref{fig:SI_Contamination}a shows examples of XPS spectra from the same sample before and after triacid cleaning. While both surfaces show identical survey scans, fine scans can reveal small Si and Na contamination peaks (cyan curves) that are removed after triacid cleaning. Typical sources of Si, Cl, Na, and other contamination include improper drying and handling, used solvent bottles, and device packaging. We have also performed similar contamination checks to develop processes such that annealing and reactive ion etching steps do not introduce surface contamination.

If the surface is contaminated before etching or annealing, micromasking and formation of surface carbides can lead to irreversible surface roughening. For example, silicon-containing polymers used in packaging, gloves, and other containers can leech onto the diamond surface, as verified by XPS. If the diamond is annealed above around 900$^\circ$C, this silicon containing contamination layer forms a carbide at the diamond surface, which cannot be removed with triacid cleaning, piranha solution, or any other acid or base that we have explored. This carbide layer then results in surface roughening through subsequent processing, such as oxygen annealing. Similarly, Na and Cl contamination is correlated with surface roughening during reactive ion etching, which we attribute to micromasking. We show in Fig.~\ref{fig:SI_RoughSurfaces} examples of surface roughness that can result from surface contamination and damage, and we discuss the consequences for electronic structure and NV coherence in Sections \ref{sec:rough_surfaces} and \ref{sec:other_samples}. 

Another source of surface contamination is adventitious carbon that is ubiquitous on surfaces exposed to atmosphere \cite{Barr_JVacSci_1995}. For NEXAFS and high-resolution XPS, we performed \textit{in situ} annealing in order to probe the intrinsic electronic structure of the surface, rather than the adventitious carbon. Figure~\ref{fig:SI_Contamination}b shows a representative NEXAFS spectrum at the carbon K-edge before and after annealing at 530$^\circ$C. After annealing, the density of unoccupied states near the conduction band edge is suppressed, indicating either the removal or rearrangment of carbon-containing groups at the diamond surface. The sample was then removed from vacuum and exposed to atmosphere for a few minutes, and re-inserted into the chamber without further annealing. Upon re-exposure to atmosphere, much of the pre-edge density returned, indicating that this pre-edge feature arises from carbon-containing contamination of the surface.

This particular NEXAFS dataset was was acquired at the NIST U7a beamline of the National Synchrotron Light Source (NSLS) at Brookhaven National Laboratory, NY.  For data taken at NSLS, the carbon K-edge and oxygen K-edge NEXAFS were collected in a partial electron yield (PEY) mode with an entrance grid bias of 220~V for carbon, and 400~V for oxygen. The incident light from a bending magnet is passed through a toroidal spherical grating monochromator, focused through a monochromator slit, and enters the chamber polarized perpendicular to the plane of incidence. The polarization with respect to the sample can then be controlled by changing the angle of the sample. Spectra are processed and calibrated by first dividing by the total incident intensity, which is measured using photocurrent from a clean gold grid in the chamber, subtracting the average pre-edge background (270--275~eV for carbon, 520--525~eV for oxygen), and normalizing to the post-edge electron yield (340--342~eV for carbon, 568--570~eV for oxygen). The energy calibration is reported relative to amorphous carbon and oxygen standards. A tantalum heater was used for \textit{in situ} annealing in the loadlock at pressures ranging from $3\times 10^{-7}$ to $8 \times 10^{-7}$ Torr.

Finally, surface contamination after processing can impact the coherence and spin relaxation of shallow NV centers. In particular, contact with metal particles from the coplanar microwave stripline can result in decreased spin relaxation time $T_1$ and coherence time $T_2$. Without the protection of the Al$_2$O$_3$ ALD layer, shallow NV centers exhibit shorter spin relaxation and coherence times when placed in microscope immersion oil on the metal stripline, as illustrated in Figure~\ref{fig:SI_Contamination}c.

\section{$sp^2$ Carbon Formation During High Temperature Annealing to Remove Etch Damage Layer}
\begin{figure}[H]
	\centering
	\includegraphics[width=0.7\columnwidth]{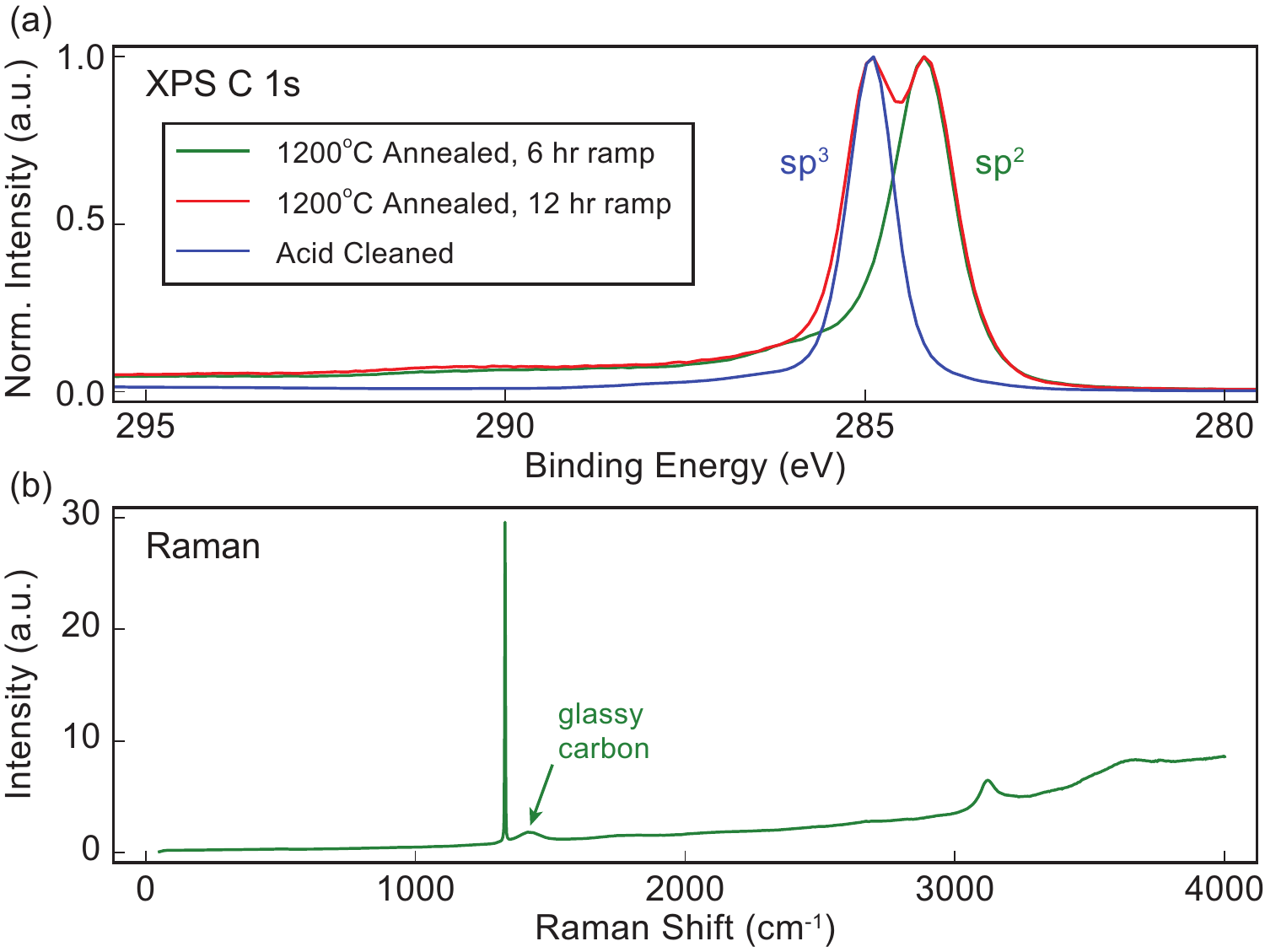}
	\caption{(a) Normalized XPS spectra of the same sample after a $1200^\circ$C anneal with different ramp rates. Material at the surface is converted to $sp^2$ carbon during annealing (red and green) and manifests as a peak at lower binding energy compared to the spectrum after the subsequent triacid clean (blue), which consists primarily of an $sp^3$ carbon peak. (b) The Raman spectrum of the 1200$^\circ$C annealed surface shows a glassy carbon peak, but no clear evidence of graphite peak at $1580~\mathrm{cm}^{-1}$ \cite{Ferrari_RoyalSociety_2004}.}
	\label{fig:SI_GlassyCarbon}
\end{figure}

XPS of the carbon $1s$ peak following the 1200$^\circ$C anneal (Fig.~\ref{fig:SI_GlassyCarbon}a) shows a significant $sp^2$ carbon peak at 284.2 eV, in addition to the diamond $sp^3$ carbon peak at 285 eV. The double-peak is consistent with an $sp^2$ carbon layer that is $<$3 nm thick, and the relative magnitude of this peak can be increased by shortening the ramp time and thus increasing the peak pressure during the vacuum anneal. Typical peak pressures are $2\times 10^{-6}$ to $5 \times 10^{-6}$ Torr for the 12 hour and 6 hour ramps, respectively. Triacid cleaning removes the $sp^2$ carbon layer, as shown by the blue curve in Fig.~\ref{fig:SI_GlassyCarbon}a. Raman spectroscopy (Horiba LabRam Evolution, 532 nm excitation)  of the surface with the $sp^2$ carbon layer shows no evidence of graphitic carbon, but instead reveals a side peak associated with glassy or amorphous carbon around $1350~\mathrm{cm}^{-1}$. 

\section{Process Calibration for Oxygen Annealing} \label{sec:O2_cal}
\begin{figure}
	\centering
	\includegraphics[width=0.7\columnwidth]{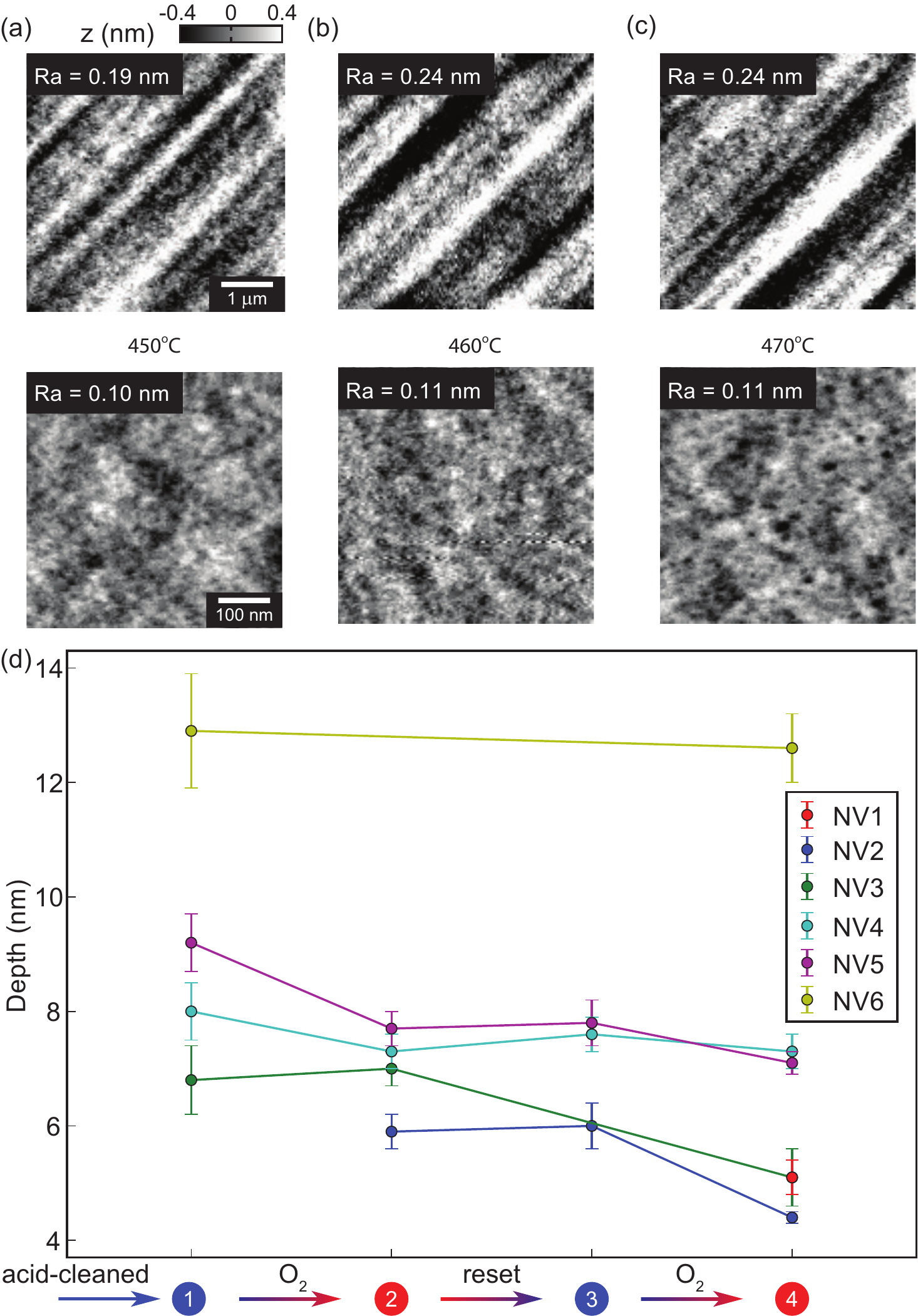}
	\caption{(a--c) AFM images of a scaife polished standard grade sample annealed at successively higher temperatures for temperature calibration of oxygen furnace. Micro-pits (black circles about 20~nm in diameter) are visible after the oxygen anneal at 470$^\circ$C. (d) Depth measurements of the six NV centers presented in the main text after different annealing steps.}
	\label{fig:SI_O2_cal}
\end{figure}

At low temperatures, oxygen annealing is not effective at changing the diamond surface chemistry, and at temperatures between 450 and 500$^\circ$C, oxygen annealing can lead to irreversible surface roughening. In order to precisely calibrate the oxygen annealing temperature, we perform detailed AFM characterization to detect the onset of surface roughening. Fig.~\ref{fig:SI_O2_cal}a--c shows AFM images from the oxygen annealing calibration after annealing at 450--470$^\circ$C in 10$^\circ$C steps. The top (bottom) row shows $5\times 5$~$\mu$m ($500\times 500$~nm) scans. The sample annealed at 470$^\circ$C shows micro-pits that are more visible in the $500\times 500$~nm scan. Therefore, we anneal the NV samples at 445--450$^\circ$C to avoid pitting.

Most failures of the oxygen anneal result in high surface roughness and prevalence of micro-pits, as shown in Fig.~\ref{fig:SI_T2_PA_MB_SK}a and Fig.~\ref{fig:SI_T2_PA_MB_SK}b. Anecdotally, we also found that after several annealing cycles with the same oxygen tank, the process starts to result in micro-pits, despite the oxygen tank being more than half full. Switching the oxygen tank more frequently results in a smooth surface with no significant change in the optimal temperature after calibration.

To monitor the etching of the diamond surface due to oxygen annealing, we measure the depths of the NV centers at each of the successive annealing steps. The results are shown in Fig.~\ref{fig:SI_O2_cal}d. While we observe a small decrease in depths (1--2~nm) on NV2, NV3, and NV5, the depths of NV4 and NV6 remain unchanged across the four processing steps. The depth decrease from the subset of NV centers could be attributed to micro-pits that occur after successive oxygen annealing (Fig.~\ref{fig:SI_O2_cal}c). For the depths presented in the main text, we use the average and standard deviation of all the depth measurements for a given NV center through different processing steps.

\section{Surface Damage from Oxygen Processing and Reactive Ion Etching} \label{sec:rough_surfaces}
\begin{figure}[H]
	\centering
	\includegraphics[width=0.7\columnwidth]{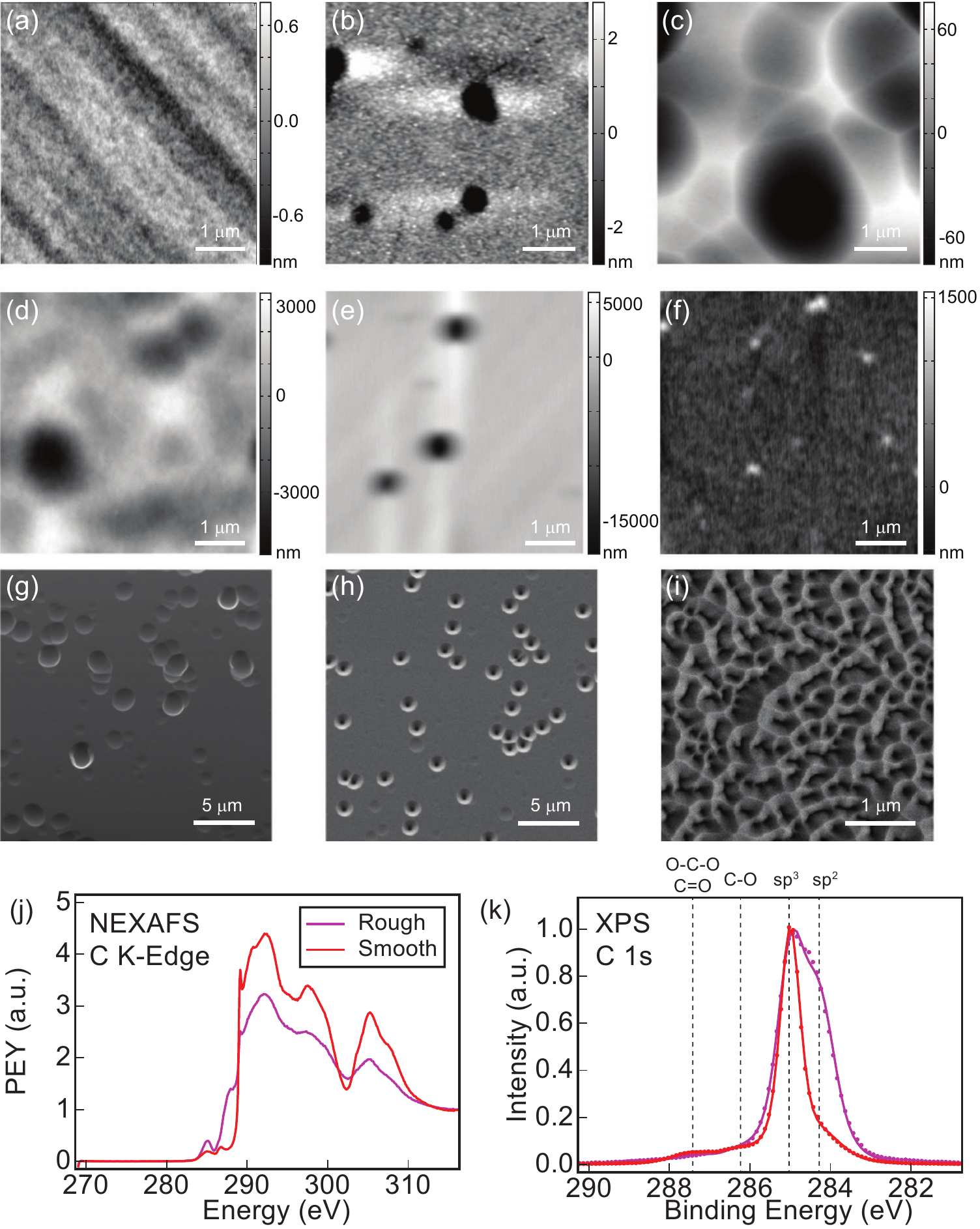}
	\caption{(a-f) Representative AFM scans showing (a) smooth surface morphology and rough surfaces resulting from (b) contamination or surface damage during an oxygen anneal of a contaminated surface, (c) RIE etching after improper polishing, (d) RIE etching of a Na and Cl contaminated surface, (e) an oxygen anneal with uncontrolled gas purity, and (f) micromasking during an oxygen etch. (g-i) SEM images of large scale morphology from a variety of damaged samples after reactive ion etching. (j) Carbon K-edge NEXAFS shows a comparison between a rough surface (purple, sample shown in (c)) and a smooth surface (red) that have been both triacid-cleaned with the full process as described in the main text. (k) High resolution carbon $1s$ XPS spectra from the two surfaces. The rough surface shows a large non-diamond $sp^2$ carbon peak at lower binding energy.}
	\label{fig:SI_RoughSurfaces}
\end{figure}

We show in Fig.~\ref{fig:SI_RoughSurfaces}b--i AFM and SEM images of various samples with rough surface morphology that result from surface contamination and subsequent processing. This surface roughness leads to drastic differences in the electronic structure at the surface, which is evident in photoelectron spectroscopy. For example, the NEXAFS and XPS spectra for the surface shown in (Fig.~\ref{fig:SI_RoughSurfaces}c) after oxygen annealing are plotted in Fig.~\ref{fig:SI_RoughSurfaces}j,k, overlaid with spectra from a smooth, oxygen-annealed surface for comparison. Fig.~\ref{fig:SI_RoughSurfaces}j shows a larger pre-edge $sp^2$ carbon peak, a much larger density of unoccupied density of states below the conduction band edge, and a lower contrast second  band gap. The carbon $1s$ XPS spectrum for this rough surface (Fig.~\ref{fig:SI_RoughSurfaces}k) also shows a significantly larger 284~eV ``non-diamond'' carbon peak. These data show that despite nominally identical processing, the rough surface exhibits very different electronic characteristics compared to a smooth surface, and thus it is important to start with and maintain the smooth surface morphology throughout various processing steps.

\section{Shallow NV Spin Coherence in Rough Samples} \label{sec:other_samples}
\begin{figure*}[h!]
	\centering
	\includegraphics[width=\columnwidth]{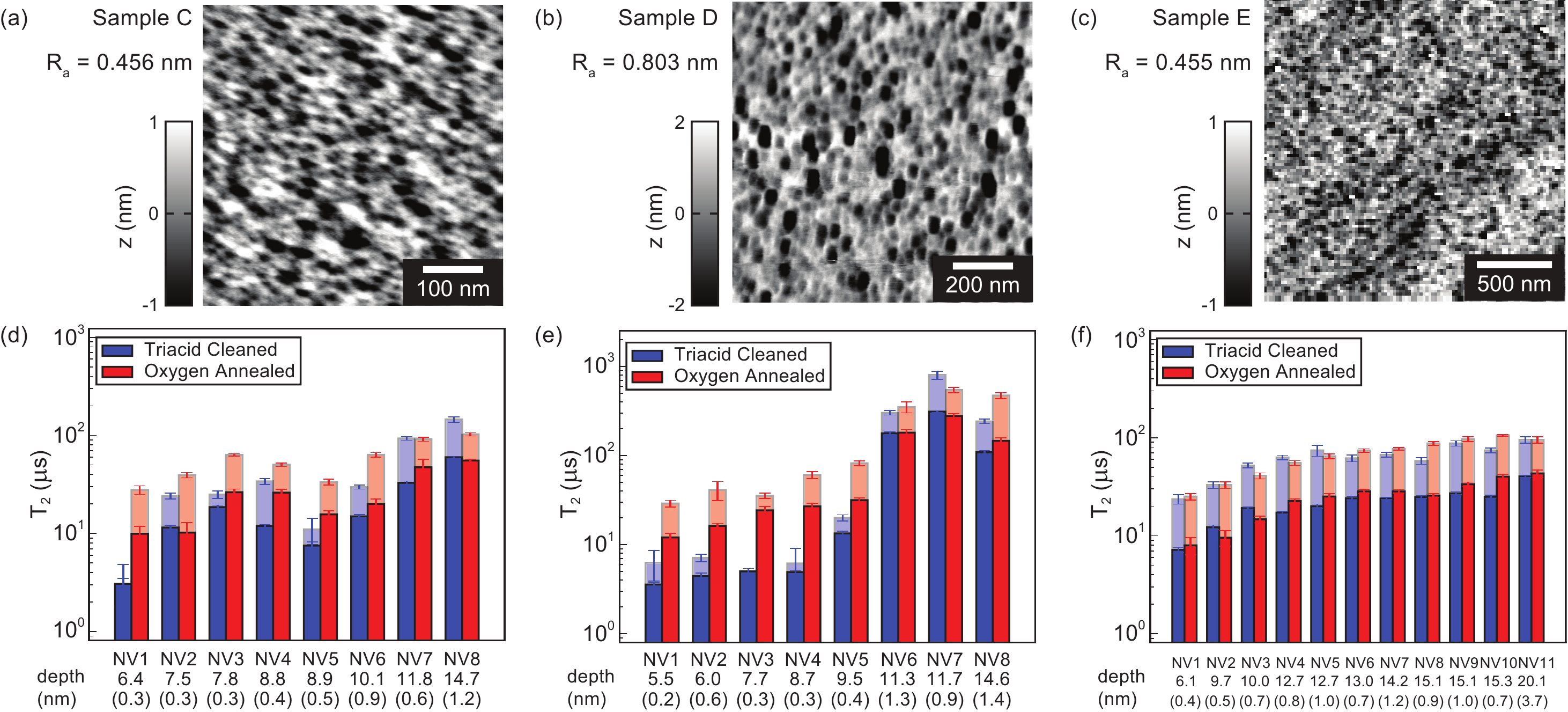}
	\caption{(a-c) AFM images and (d-f) NV coherence times $T_{2,\mathrm{Echo}}$ (dark) and $T_{2, \mathrm{XY8}}$ (light) from three additional samples with rougher surface morphologies. The improvement in coherence time after oxygen annealing is less pronounced than that of Sample~A.}
	\label{fig:SI_T2_PA_MB_SK}
\end{figure*}

In addition to the data in the main text, we also performed oxygen annealing on three additional samples: Samples~C, D, and E (Fig.~\ref{fig:SI_T2_PA_MB_SK}). While we observe improvements in coherence times for Sample~C and Sample~D after oxygen annealing, the effects are not as pronounced as that of Sample~A. Moreover, Sample~E shows essentially no improvement.

By examining the surface properties of these samples, we found that the surfaces are morphologically different from Sample~A. These samples had all undergone extensive prior processing. Notably, all three samples have higher average roughness and show micropits that we do not observe in Sample~A. These results suggest that subtle differences in surface morphology have a significant effect on NV coherence times, even after oxygen annealing.

\section{Dynamical Decoupling Data}
\begin{figure}[H]
	\centering
	\includegraphics[width=0.7\columnwidth]{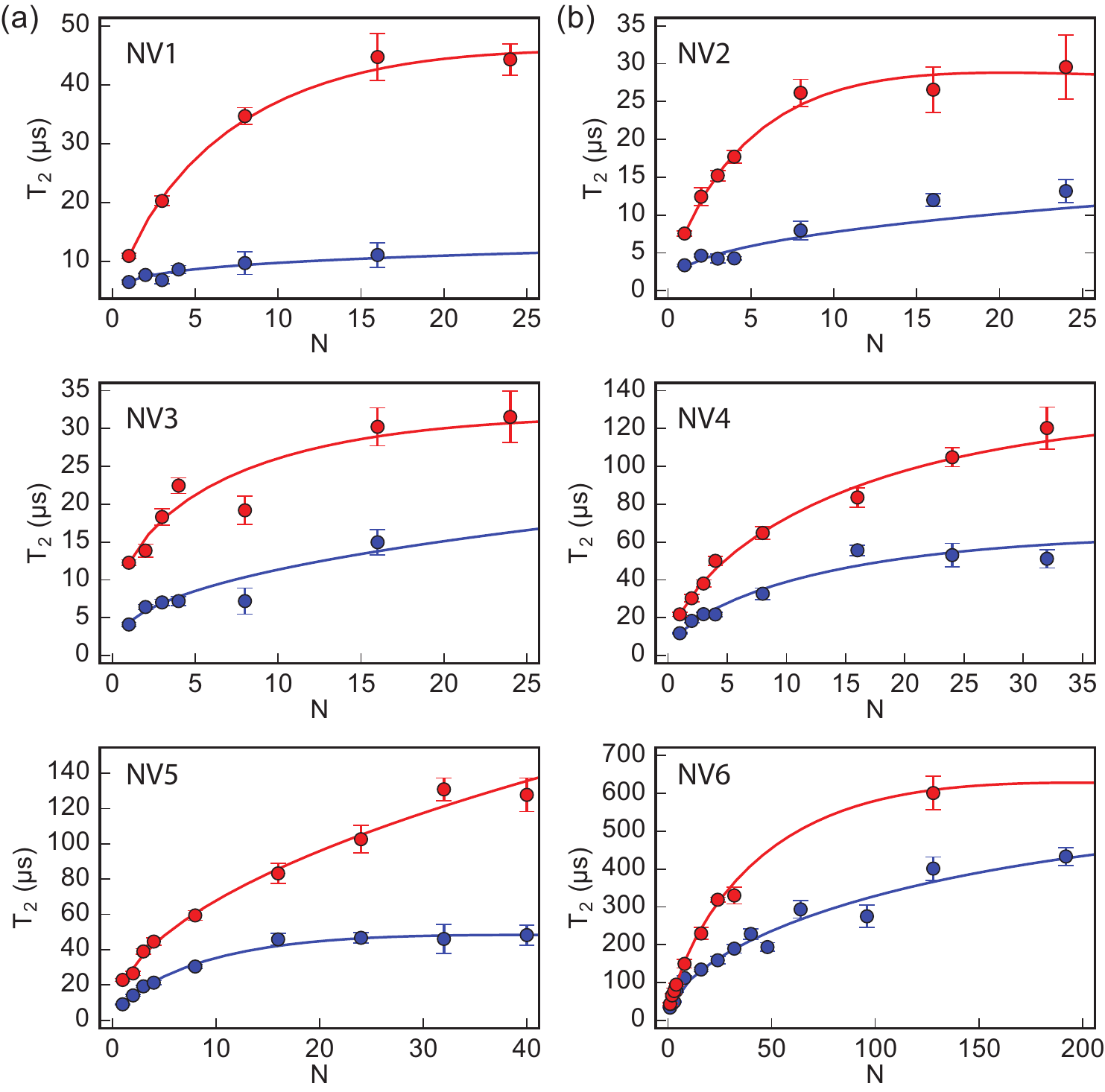}
	\caption{Coherence time as a function of number of dynamical decoupling pulses, $N$, for all six NV centers presented in the main text. The fit parameters are shown in Table~\ref{tab:SI_T2_Scaling}.}
	\label{fig:SI_T2_Scaling}
\end{figure}

We show in Fig.~2b that shallow NV centers under both triacid-cleaned and oxygen-annealed surfaces show scaling of coherence time with number of pulses that differ from that of a slowly fluctuating spin bath, $T_2 \propto N^{2/3}$. We fit the data to a saturation curve, given by
\begin{equation}
T_2 (N) = T_2(1) [N_\mathrm{sat}^s + (N^s - N_\mathrm{sat}^s) e^{N/N_\mathrm{sat}}],
\end{equation}
or to a power law, given by
\begin{equation}
T_2 (N) = T_2(1) N^s,
\end{equation}
when there is no observable saturation of $T_2$ \cite{Romach_PRL_2015}. The fitted curves are shown in Fig.~\ref{fig:SI_T2_Scaling} and the fitting parameters for the two surface terminations are shown in Table~\ref{tab:SI_T2_Scaling}.
\begin{table*}
	\begin{tabular}{|c|c|c|c|c|c|c|c|c|c|}
		\hline
		\multirow{2}{*}{NV} & \multirow{2}{*}{d [nm]} & \multicolumn{4}{c|}{Triacid-cleaned} & \multicolumn{4}{c|}{oxygen-annealed} \\
		\cline{3-10}
		& & $s$ & $T_2(1)\ [\mathrm{\mu s}]$ & $N_\mathrm{sat}$ & $T_{2,\mathrm{sat}}\ [\mathrm{\mu s}]$ & $s$ & $T_2(1)\ [\mathrm{\mu s}]$ & $N_\mathrm{sat}$ & $T_{2,\mathrm{sat}}\ [\mathrm{\mu s}]$\\
		\hline
		1 & 5.1(3) & 0.17(5) & 6.6(3) & - & - & 0.56(4) & 8.8(5) & 16(2) & 43.0(6.4) \\
		2 & 5.5(4) & 0.39(9) & 3.1(4) & - & - & 0.66(5) & 5.7(4) & 10(1) & 26.7(4.5) \\
		3 & 5.8(9) & 0.41(6) & 4.4(3) & - & - & 0.32(8) & 11.3(1.3) & 22(27) & 30.5 (15.4) \\
		4 & 7.2(4) & 0.49(7) & 10.4(1.4) & 35(27) & 60(31) & 0.49(4) & 19.4(1.5) & 41(19) & 120(36) \\
		5 & 7.5(3) & 0.60(4) & 7.4(4) & 20(3) & 45.2(7.6) & 0.50(2) & 21.4(9) & - & - \\
		6 & 12.1(9) & 0.47(2) & 31.5(2.1) & 388(341) & 517(248) & 0.66(4) & 30.2(3.5) & 51(9) & 586(133) \\
		\hline
	\end{tabular}
	\caption{Fit parameters for the dynamical decoupling scaling shown in Fig.~\ref{fig:SI_T2_Scaling}.}
	\label{tab:SI_T2_Scaling}
\end{table*}

\section{Noise Spectral Density Extracted from NV Center Measurements}
\subsection{Dynamical Decoupling Spectral Decomposition}
\begin{figure}[H]
	\centering
	\includegraphics[width=0.7\columnwidth]{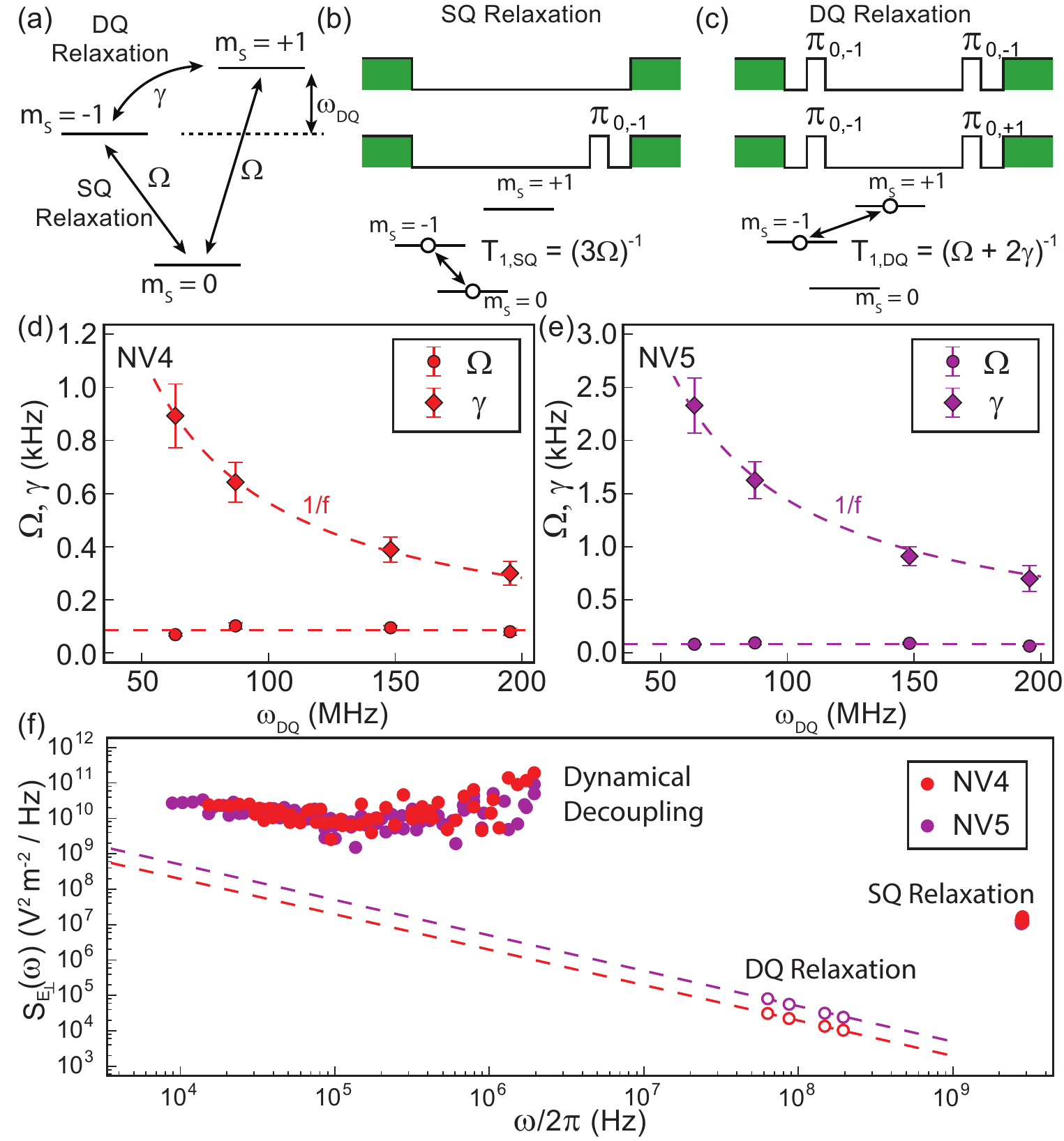}
	\caption{(a) Energy level diagram of the ground state electronic spin of the NV center, indicating single- and double-quantum relaxation channels, whose rates are denoted by $\Omega$ and $\gamma$, respectively. (b) SQ and (c) DQ relaxation measurement sequences. The NV is prepared in the $m_S = 0$ ($m_S = -1$) state and the relaxation to $m_S = -1$ ($m_S = +1$) is measured. (d,e) SQ and DQ transitionrates as a function of DQ transition frequency $\omega_\mathrm{DQ}$ for two different NV centers. Dashed lines indicate the broadband spectra for $\Omega$ and $1/f$ spectra for $\gamma$. (f) Comparison of the noise spectral density from single- and double-quantum measurements. The two spectra are plotted on the same scale using the effective electric field noise $S_{E_\perp}$. We found that magnetic field noise (solid circles) is the dominant source of decoherence for oxygen-annealed surfaces, compared to the electric field noise (open circles).}
	\label{fig:SI_SpectralDecomp}
\end{figure}

We probe the spectral density of the magnetic field noise bath from the oxygen-terminated surface using dynamical decoupling. From the coherence decay $C(T)$, where $T=N\tau$ is the total free precession time, the noise spectrum $S(\omega)$ can be obtained using Eq.~\ref{eq:spectral_1} and Eq.~\ref{eq:spectral_2}.
\begin{align}
C(T) &= \exp [-\chi(T)] \label{eq:spectral_1}\\
\chi(T) = -\ln C(T) &= \frac{1}{\pi} \int_0^\infty S(\omega) \frac{F_N(\omega T)}{\omega^2} \approx \frac{TS(\omega)}{\pi} \label{eq:spectral_2}
\end{align}
Spectral decomposition reveals a broadband noise spectrum across the frequency range of 0.01--1~MHz (Fig.~\ref{fig:SI_SpectralDecomp}f).

\subsection{Double Quantum Relaxometry}
To probe the electric field noise, we follow the procedure and notation from Myers~\textit{et~al.} \cite{Myers_PRL_2017}. We briefly outline their methods here, starting from the NV Hamiltonian, given by
\begin{equation}
H/h = (D+d_\parallel\Pi_\parallel)S_z^2 + g\mu_B \vec{B}\cdot \vec{S} -  \frac{d_\perp\Pi_\perp}{2}(S_+^2 + S_-^2), \label{eq:Hamiltonian}
\end{equation}
where $D = 2.87 \mathrm{\ GHz}$ is the zero-field splitting, $g\mu_B = 2.8 \mathrm{\ MHz/G}$ is the electron gyromagnetic ratio, and $d_\parallel = 0.35 \mathrm{\ Hz\cdot cm/V}, d_\perp = 17 \mathrm{\ Hz\cdot cm/V}$ are the electric dipole moments. From the Hamiltonian, we can see that while the magnetic field $\vec{B}$ can only drive the single-quantum (SQ) transition ($\Delta m_s = \pm 1$), the electric field component $\Pi_\perp$ can result in a double-quantum (DQ) transition ($\Delta m_s = \pm 2$) from the last term in Eq.~\ref{eq:Hamiltonian}. Therefore, a full three-level model is required to account for the full spin relaxation resulting from magnetic and electric field noise. The three-level energy diagram is depicted in Fig.~\ref{fig:SI_SpectralDecomp}a, with the SQ and DQ transition rates denoted by $\Omega$ and $\gamma$, respectively. The pulse sequences for measuring the relaxation times $T_{1,\mathrm{SQ}}$ and $T_{1,\mathrm{DQ}}$ are shown in Fig.~\ref{fig:SI_SpectralDecomp}b,c.
From SQ and DQ $T_1$ measurements, we can extract the SQ and DQ transition rates, $\Omega$ and $\gamma$, from Eq.~\ref{eq:sqdq_t1} \cite{Myers_PRL_2017}.
\begin{equation}
T_{1, \mathrm{SQ}} = (3\Omega)^{-1}, \hspace{0.5cm} T_{1, \mathrm{DQ}} = (\Omega + 2\gamma)^{-1}\label{eq:sqdq_t1}
\end{equation}
The electric field noise spectrum can be obtained directly from the DQ transition rate $S_\gamma(\omega_\mathrm{DQ}) = \gamma(\omega_\mathrm{DQ})$, where $\omega_\mathrm{DQ}$ is the double-quantum transition frequency. Figure~\ref{fig:SI_SpectralDecomp}d,e shows the noise spectra sampled from NV4 and NV5, consistent with $1/f$ noise. The magnetic field noise spectrum can be obtained from the SQ transition rate $S_\Omega(\omega_\mathrm{SQ}) = \Omega(\omega_\mathrm{SQ})$, where $\omega_\mathrm{SQ} = D - g\mu_B B_z = D - \omega_\mathrm{DQ}/2$. Unlike $S_\gamma$, $S_\Omega$ is roughly constant across this magnetic field range.
\subsection{Comparing Magnetic and Electric Noise Spectra}
Even though the magnetic field noise spectrum obtained from SQ measurements (dynamical decoupling and SQ relaxation) and the electric field noise spectrum obtained from DQ measurements are of distinct origins, we can place them on the same scale for comparison by considering the effective field $E_\perp$ \cite{Myers_PRL_2017}, given by
\begin{equation}
S_{E_\perp}^\mathrm{SQ} (\omega) = 2S_{E_\parallel}^\mathrm{SQ} (\omega) = 2 \frac{S_\mathrm{SQ} (\omega)}{d_\parallel^2/h^2}, \hspace{0.5cm}
S_{E_\perp}^\mathrm{DQ} (\omega) = \frac{S_\mathrm{DQ} (\omega)}{d_\perp^2/h^2},
\end{equation}
where $S_{E_\parallel}^\mathrm{DD}$ is obtained by considering an effective electric field that would cause first-order dephasing and the relationship $\langle E_\perp \rangle ^2 = 2 \langle E_\parallel \rangle ^2$ is obtained by considering the geometry of the NV axis on a (100)-oriented diamond. The combined spectrum is shown in Fig.~\ref{fig:SI_SpectralDecomp}f. By extrapolating the electric field noise, obtained from the DQ spectra (dashed lines), we found the amplitude of the magnetic field noise, obtained from dynamical decoupling and SQ relaxation, is the dominant source of noise for the oxygen-annealed surface.

\begin{figure}[h!]
	\centering
	\includegraphics[width=0.7\columnwidth]{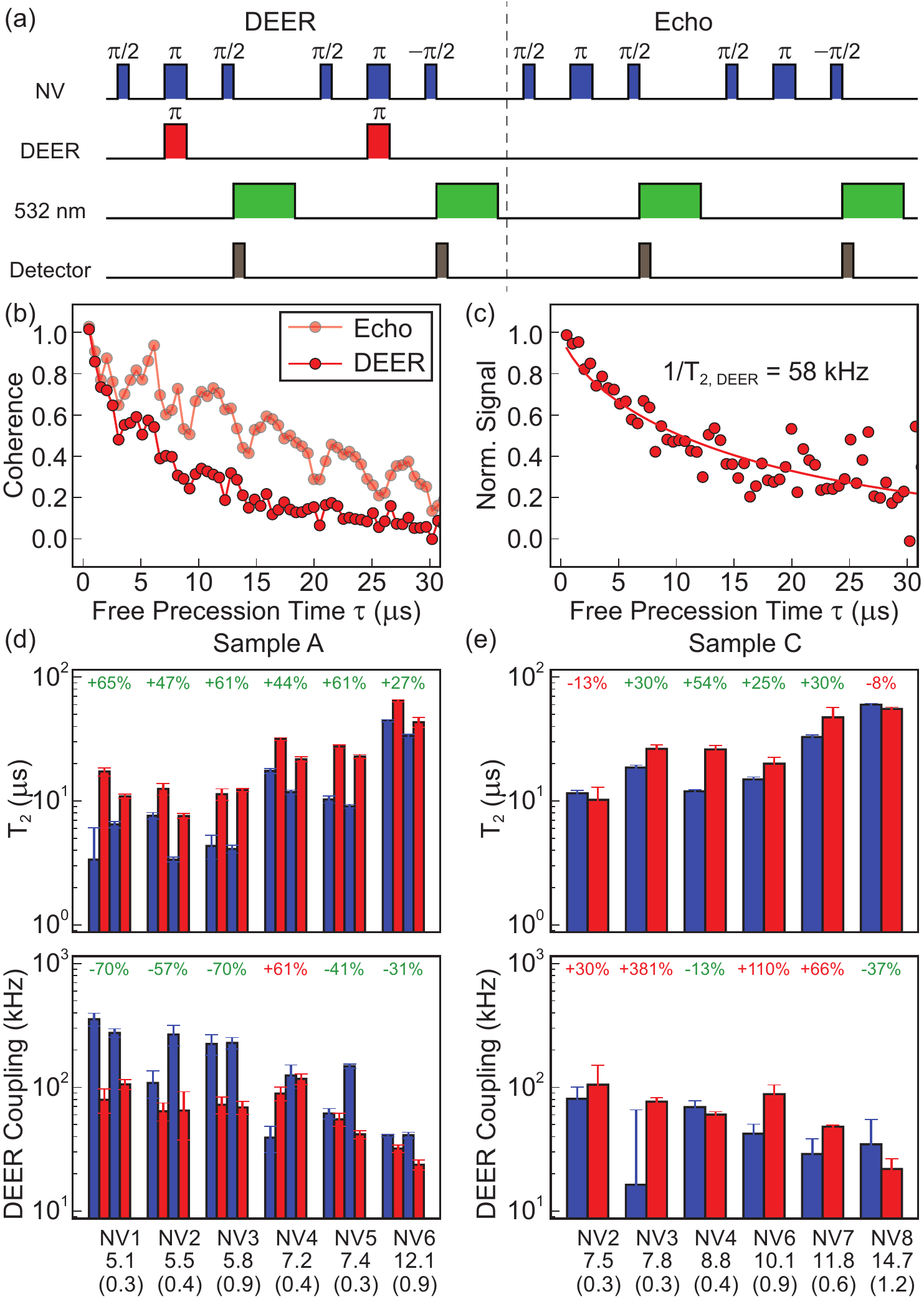}
	\caption{(a) Pulse sequence for measuring DEER coupling: a DEER sequence is interleaved with a Hahn echo sequence to mitigate effects from slow system drift. (b) Hahn echo and DEER decay curves taken at $B_z \approx 330 \mathrm{\ G}$. Collapse and revival arising from hyperfine coupling with a nearby $^{13}$C can be observed. (c) Normalized signal, where the $^{13}$C modulation is no longer pronounced. (d-e) Spin echo coherence time and DEER coupling for several NV centers in Sample~A and Sample~C before and after oxygen annealing. Green (red) numbers indicate the average improvement (worsening) in the decoherence rate $1/T_2$, and the average reduction (increase) in the DEER coupling after oxygen annealing.}
	\label{fig:SI_DEER}
\end{figure}

\subsection{Coupling to Dark Surface Spins}
Several experiments have reported that the coherence times of the shallow NV centers are limited by magnetic noise arising from ``dark" surface spins that are not optically active \cite{Lovchinsky_Science_2016, Myers_PRL_2014, FavarodeOliveira_NatComm_2017, Rosskopf_PRL_2014}. To examine this hypothesis, we probe the coupling of our NV centers to dark spins via double electron-electron resonance (DEER) spectroscopy, where a second microwave tone is applied to flip the dark spins during the NV $\pi$-pulse in the Hahn echo sequence. To extract the DEER coupling, we model the decoherence of the NV center with two separable sources of decoherence, one arising from the dark spins, and another decoherence rate attributable to everything else. The coherence $C$ as a function of free precession time $t$ can be written as:
\begin{align}
C_\mathrm{NV}(t) &= \exp \left[ - \left( \frac{t}{T_\mathrm{2,NV}} \right)^{N_\mathrm{NV}} \right]
\\
C_\mathrm{DEER}(t) &= \exp \left[ - \left(\left( \frac{t}{T_\mathrm{2,DEER}} \right)^{N_\mathrm{DEER}} + \left( \frac{t}{T_\mathrm{2,NV}} \right)^{N_\mathrm{NV}}\right)\right]
\end{align}
To directly measure the DEER coupling, we can normalize the DEER decay curve using the Hahn echo decay curve. The resulting decay curve can be used to extract DEER coupling $g_\mathrm{DEER} = 1/T_{2,\mathrm{DEER}}$.
\begin{align}
\frac{C_\mathrm{DEER}(t)}{C_\mathrm{NV}(t)} &= \exp \left[ - \left( \frac{t}{T_\mathrm{2,DEER}} \right)^{N_\mathrm{DEER}} \right]
\end{align}
Experimentally, we interleave the DEER experiment with the Hahn echo experiment in order to mitigate any long term drift. The pulse sequence and a sample set of data are shown in Fig.~\ref{fig:SI_DEER}a--c. We performed DEER measurements on the same NV centers across several surface processing steps.
We observe that for Sample~A, whose coherence data is presented in the main text, the oxygen annealing also results in a lower DEER coupling across all the NV centers, and this lower DEER coupling is reversible and reproducible. In section~\ref{sec:other_samples}, we show that across several samples with different surface morphology, there are varying degrees of improvement to the coherence times of the NV centers with oxygen annealing, but the oxygen-annealed surface shows consistently better coherence. However, the change in DEER coupling is inconsistent among these samples, suggesting that rough surface morphology can lead to a population of persistent dark spins that are not eliminated by oxygen annealing. Figure~\ref{fig:SI_DEER}d--e shows a comparison of changes in $T_2$ and DEER coupling between Sample~A and Sample~C. While the oxygen annealing improves the coherence times in both samples and reduces the DEER coupling in Sample~A, the DEER coupling stays the same or increases in Sample~C.

We calculate the Pearson's product-moment correlation coefficient between the coherence time improvement and DEER coupling reduction for both samples and we find no significant correlation (correlation coefficients $-0.44$ and $+0.25$, respectively).

\section{Electron Affinity Measurement}
\begin{figure}[H]
	\centering
	\includegraphics[width=0.7\columnwidth]{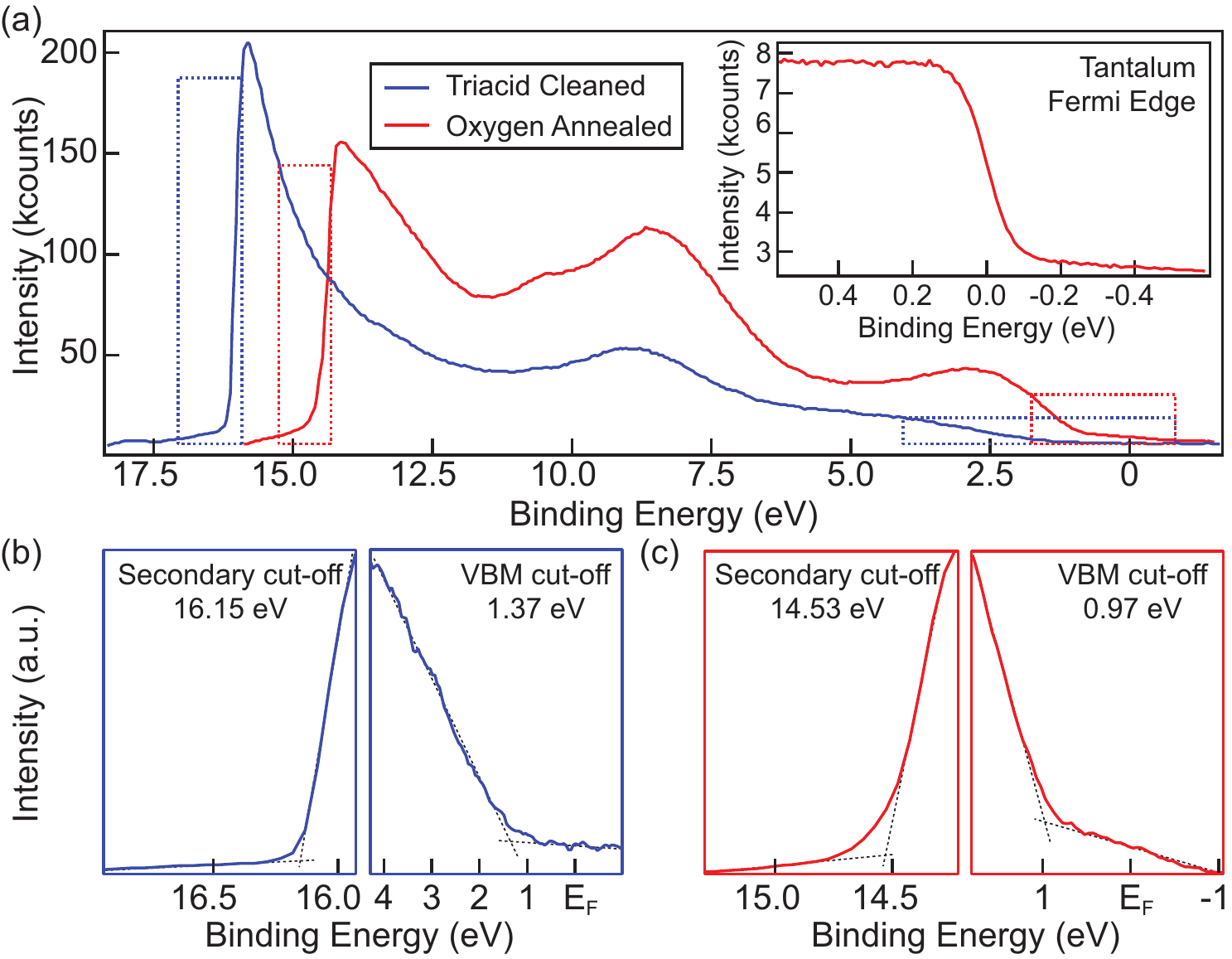}
	\caption{(a) UPS spectra from triacid-cleaned sample (blue) and oxygen-annealed sample (red) as a function of binding energy relative to the Fermi level. The Fermi level is calibrated with a reference tantalum foil (inset). The spectral width, $\omega $, is calculated as the energy difference between the secondary electron cut-off and valence band maximum, indicated by the dashed rectangles. (b,c) Zoom-in of the secondary electron cut-off and the valence band maximum for the triacid-cleaned sample and the oxygen-annealed sample, respectively.}
	\label{fig:SI_UPS}
\end{figure}

The electron affinity was calculated from the measured spectral width in UPS (Fig.~3d, also duplicated in Fig.~\ref{fig:SI_UPS}a). The binding energy scale was calibrated to the core levels and Fermi edge of a tantalum calibration sample (Fig.~\ref{fig:SI_UPS}a, inset). The width of the electron distribution curve, $\omega$, can be measured as the energy difference between the valence band maximum (VBM) and the secondary electron cutoff at high binding energy, as shown in Fig.~\ref{fig:SI_UPS}b,c. The ionization energy, $E_i$, can then be used to relate the spectral width, $\omega$, to the electron affinity, $E_A$, by
\begin{equation}
E_i = h\nu - \omega = E_g + E_A, \label{eq:UPS}
\end{equation}
where $h\nu = 21.2$~eV is the excitation energy from He I UV source and $E_g = 5.5$~eV is the band gap of the diamond.

\section{LEED Pattern of Oxygen-annealed Surface}
\begin{figure}[H]
	\centering
	\includegraphics[width=0.7\columnwidth]{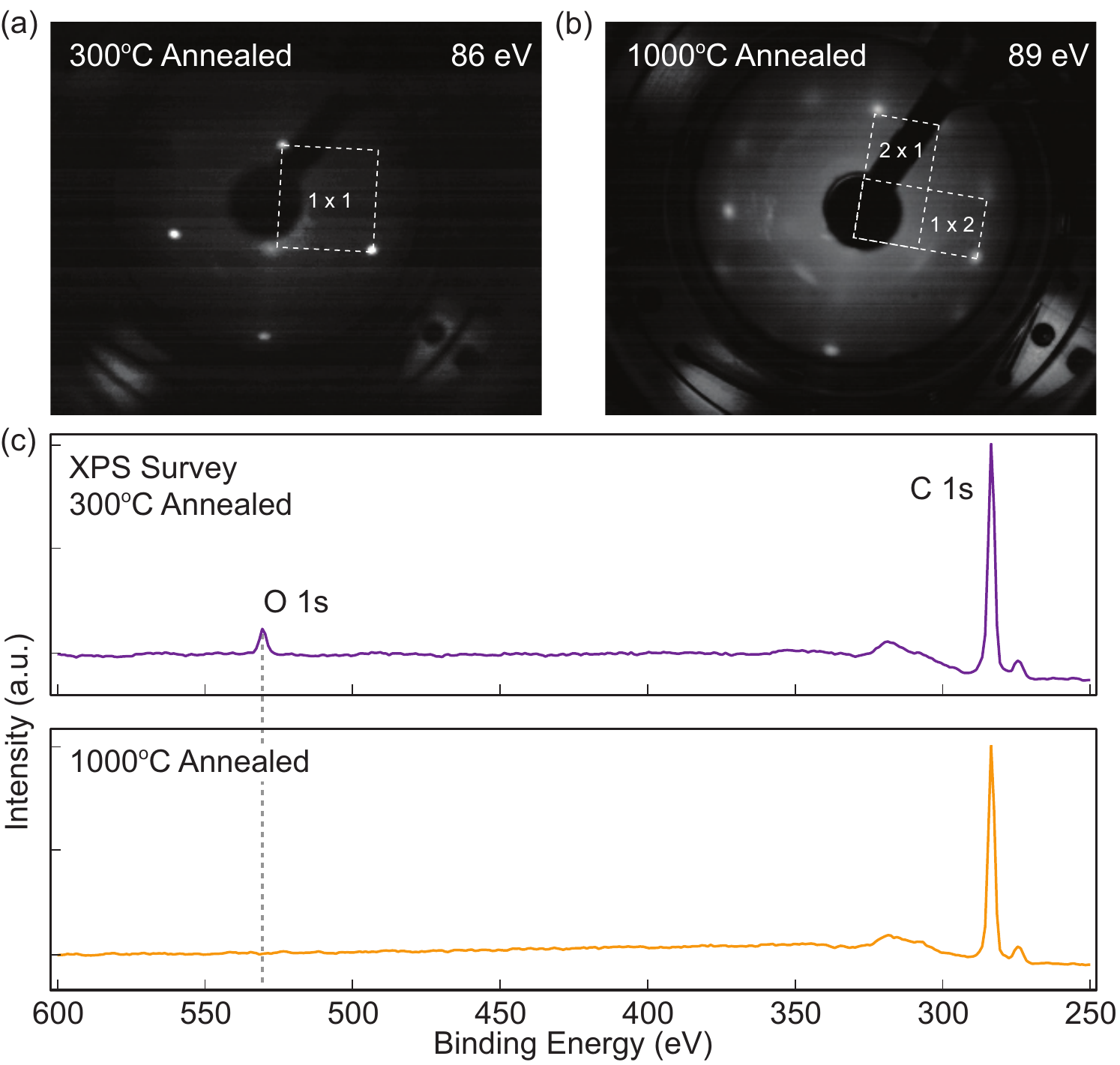}
	\caption{(a) LEED of an oxygen-annealed surface, as shown in Fig.~3c. The $1\times1$ reconstruction is consistent with an ether-terminated surface. (b) LEED after \textit{in situ} annealing at 1000$^\circ$C showing $2\times1$ reconstruction peaks that appear for the same surface after the surface heteroatoms are removed. (c) XPS survey data showing that the O $1s$ peak indeed disappears after annealing at 1000$^\circ$C.}
	\label{fig:SI_LEED}
\end{figure}

The LEED pattern for the oxygen-annealed surface (Fig.~3c, duplicated in Fig.~\ref{fig:SI_LEED}a) shows that the surface is $1\times1$ reconstructed, and there is no evidence of $2\times1$ reconstruction. In order to verify that the absence of $2\times1$ diffraction peaks does not arise from disorder obscuring the peaks, the sample was annealed at 1000$^\circ$C to remove the oxygen termination, as verified by XPS (Fig.~\ref{fig:SI_LEED}c). A clear sign of $2\times1$ reconstruction emerges in the diffraction spots (Fig.~\ref{fig:SI_LEED}b), which are absent before annealing (Fig.~\ref{fig:SI_LEED}a).

\section{Comparison of XPS Spectra for Triacid-cleaned and Oxygen-annealed Surfaces}
\begin{figure}[H]
	\centering
	\includegraphics[width=0.7\columnwidth]{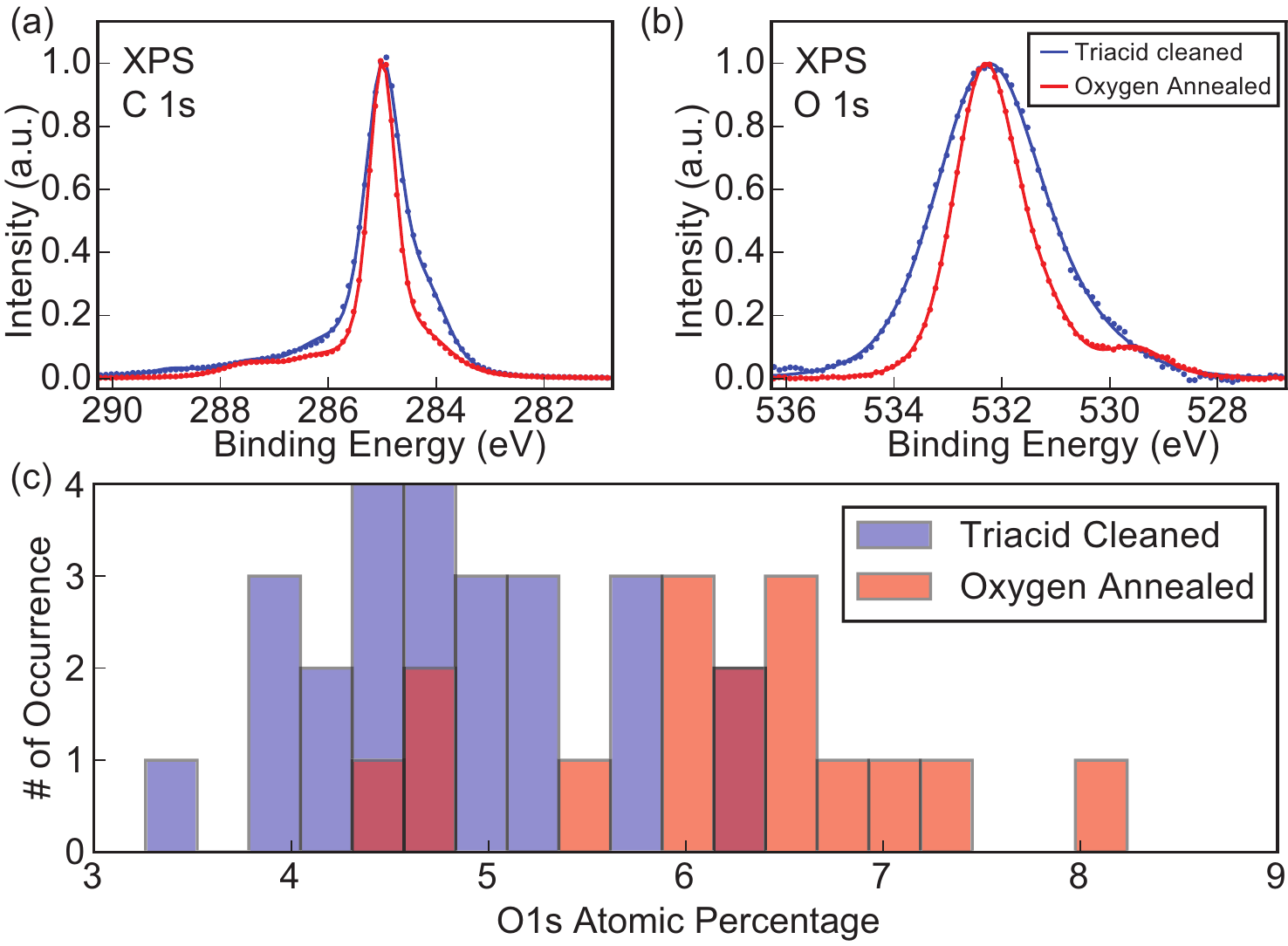}
	\caption{(a, b) Comparison of high-resolution XPS spectra at carbon $1s$ and oxygen $1s$ peaks from triacid-cleaned and oxygen-annealed surfaces. The linewidths are narrower for the oxygen-annealed surface. (c) Histogram of the oxygen $1s$ atomic percentage from the XPS signal, obtained from a large number of triacid-cleaned and oxygen-annealed surfaces.}
	\label{fig:SI_XPS_acid_vs_oxygen}
\end{figure}

High-resolution XPS spectra were measured at the Australian Synchrotron. The carbon $1s$ spectrum (Fig.~\ref{fig:SI_XPS_acid_vs_oxygen}a) for the triacid-cleaned sample shows a larger peak width compared to the oxygen-annealed surface, individual satellite peaks cannot be resolved, and the $sp^2$ carbon peak at lower binding energy is more pronounced. There is also some weight to the spectrum at higher binding energies, above 288~eV, possibly indicating the presence of some carboxylic acid groups at the surface \cite{Strobel_DiamRelMat_2008}. The oxygen $1s$ spectrum (Fig.~\ref{fig:SI_XPS_acid_vs_oxygen}b) for the triacid-cleaned surface also shows a single dominant peak with some species at lower binding energies, but the peak width is 2.1~eV, compared to 1.3~eV for the oxygen-annealed surface. The broader peak widths are consistent with a highly disordered, heterogeneous surface termination.

XPS survey scans of many samples across many processing steps were taken at the Princeton IAC. From the survey scans, the atomic composition of the surface can also be estimated by comparing the magnitude of the XPS peaks. Here, we focus on the oxygen $1s$ spectra from triacid-cleaned surfaces and oxygen-annealed surfaces. From the inelastic mean free path of photoelectrons at these energies (2.2~nm for 1487~eV Al K$\alpha$) \cite{Shinotsuka_SurfIntAnalysis_2015}, we estimate the contribution of the signal from a monolayer of atoms on the diamond surface to be $7.6\%$. Figure~\ref{fig:SI_XPS_acid_vs_oxygen}c shows a histogram of the measured oxygen $1s$ atomic percentage from multiple XPS spectra, where the oxygen-annealed surfaces show markedly higher oxygen $1s$ percentage compared to the triacid-cleaned surfaces. The values obtained are consistent with an oxygen monolayer on the surface after oxygen annealing, while the triacid-cleaned surfaces have lower oxygen coverage, indicating incomplete oxygen termination.

\bibliography{SI} 